\documentclass[]{aa}  
\usepackage[varg]{txfonts}
\usepackage{natbib}
\usepackage{lipsum}
\usepackage{rotating}
\usepackage[breaklinks=true,colorlinks=true,linkcolor=blue,citecolor=blue]{hyperref}
\usepackage{placeins}
\bibpunct{(}{)}{;}{a}{}{,}
\usepackage[usenames,dvipsnames,svgnames,table]{xcolor}
\definecolor{violet}{HTML}{9400D3}

\newcommand{\li}[3]{\ion{#1}{#2}~$\lambda$#3\,$\angstrom$}
\newcommand{\Teff}{\ensuremath{T_\mathrm{eff}}}
\newcommand{\msun}{\ensuremath{\mathrm{M}_\sun}}
\newcommand{\lsun}{\ensuremath{\mathrm{L}_\sun}}

\newcommand{\rsun}{\ensuremath{\mathrm{R}_\sun}}
\newcommand{\Zsun}{\ensuremath{Z_\sun}}
\newcommand{\Zini}{\ensuremath{Z_\mathrm{ini}}}
\newcommand{\rstar}{\ensuremath{R_*}}
\newcommand{\tstar}{\ensuremath{T_*}}
\newcommand{\mstar}{\ensuremath{M_*}}
\newcommand{\lstar}{\ensuremath{L_*}}

\newcommand{\msunyr}{{\ensuremath{\msun}/\mathrm{yr}}}
\newcommand{\vel}{{v}}

\newcommand{\vinfty}{\ensuremath{\vel_\infty}}
\newcommand{\vcrit}{\ensuremath{\vel_\mathrm{crit}}}
\newcommand{\vesc}{\ensuremath{\vel_\mathrm{esc}}}
\newcommand{\vinf}{\ensuremath{\vel_\infty}}
\newcommand{\vdop}{\ensuremath{\vel_\mathrm{D}}}

\newcommand{\mdot}{\ensuremath{\dot M}}
\newcommand{\second}{\mbox{s}}
\newcommand{\kms}{\ensuremath{\mbox{km}/\second}}
\newcommand{\angstrom}{\text{\normalfont\AA}}
\newcommand{\cmss}{\ensuremath{\mbox{cm}/\second^{2}}}
\newcommand{\Kelvin}{\ensuremath{\mbox{K}}}
\newcommand{\kKelvin}{\ensuremath{\mbox{kK}}}
\newcommand{\microm}{\ensuremath{\mu\mbox{m}}}
\newcommand{\PoWR}{{\tt PoWR}}
\newcommand{\Halpha}{\ensuremath{\mathrm{H}\alpha}}
\newcommand{\YC}{\ensuremath{Y_\mathrm{C}}}
\newcommand{\YS}{\ensuremath{Y_\mathrm{S}}}
\newcommand{\mini}{\ensuremath{M_\mathrm{ini}}}
\newcommand{\zav}[1]{\left(#1\right)}
\newcommand{\hzav}[1]{\left[#1\right]}
\defcitealias{Szecsi:2015}{Paper~I}

\begin{document}

	\title{Low-metallicity massive single stars with rotation}
%	\subtitle{II. Predicting spectra and spectral classes of TWUIN~stars}
\subtitle{II. Predicting spectra and spectral classes of chemically-homogeneously evolving stars}
%     \titlerunning{\bf Low-metallicity massive stars. Part II. TWUIN spectra}
     \titlerunning{Low-metallicity massive stars. Part II. Spectra of chem.~hom. evolving stars}

	\author{B. Kub\'atov\'a
		\inst{\ref{inst1}}
		\and
		D. Sz\'ecsi\inst{\ref{inst1},\ref{inst2}}
		\and
		A.A.C. Sander\inst{\ref{inst3},\ref{inst4}}
		\and
		J. Kub\'at\inst{\ref{inst1}}
		\and
		F. Tramper\inst{\ref{inst5},\ref{inst9}}
		\and
		J. Krti\v{c}ka\inst{\ref{inst6}}
		\and
		C. Kehrig\inst{\ref{inst7}}
		\and
		W.-R. Hamann\inst{\ref{inst3}}
		\and
		{R. Hainich\inst{\ref{inst3}}}
		\and
		{T. Shenar\inst{\ref{inst8}}}
	}
	
	\institute{Astronomick\'y \'ustav, Akademie v\v{e}d \v{C}esk\'e republiky,
		Fri\v{c}ova 298, 251~65 Ond\v{r}ejov, Czech Republic
		\email{brankica.kubatova@asu.cas.cz}\label{inst1}
		\and
		School of Physics and Astronomy and Institute of Gravitational Wave Astronomy, University of Birmingham, Edgbaston, Birmingham B15 2TT, UK\label{inst2}
		\and
		Institut f\"ur Physik und Astronomie, Universit\"at Potsdam, Karl-Liebknecht-Str. 24/25, 14476, Potsdam, Germany\label{inst3}  
		\and 
		Armagh Observatory and Planetarium, College Hill, Armagh,
		BT61 9DG, Northern Ireland\label{inst4}
		\and
		European Space Astronomy Centre (ESA/ESAC), Operations Department, 28692 Villanueva de la Ca\~{n}ada, Madrid, Spain\label{inst5}
		\and
		\'{U}stav teoretick\'e fyziky a astrofyziky, Masarykova univerzita, Kotl\'a\v rsk\'a 267/2, 611 37, Brno, Czech Republic\label{inst6}
		\and
		Instituto de Astrof\'isica de Andaluc\'ia (IAA/CSIC), Glorieta de la Astronom\'ia s/n Aptdo. 3004, E-18080 Granada, Spain\label{inst7}  
		\and
		{Institute of Astrophysics, KU Leuven, Celestijnenlaan 200 D, 3001 Leuven, Belgium\label{inst8} }
		\and
		{Institute for Astronomy, Astrophysics, Space Applications \& Remote Sensing, National Observatory of Athens, Vas. Pavlou and I. Metaxa, Penteli 15236, Greece\label{inst9}}
	}
	
	\date{Received 02/10/2018; accepted ...}
	
	\abstract{Metal-poor massive stars are supposed to be progenitors of certain supernovae, gamma-ray bursts and compact object mergers, potentially contributing to the early epochs of the Universe with their strong ionizing radiation. However, they remain mainly theoretical as individual spectroscopic observations of such objects have rarely been carried out below the metallicity of the Small Magellanic Cloud.}
	{This work aims at exploring what our state-of-the-art theories of stellar evolution combined with those of stellar atmospheres predict about a certain type of metal-poor (0.02~Z$_{\odot}$) hot massive stars, the chemically homogeneously evolving ones, called TWUIN~stars.}
	{Synthetic spectra corresponding to a broad range in masses (20$-$130~M$_{\odot}$) and covering several evolutionary phases from the zero-age main-sequence up to the core helium-burning stage were computed. The influence of mass loss and wind clumping on spectral appearance were investigated, and the spectra were classified according to the MK~spectral types.}
	{We find that TWUIN~stars show almost no emission lines during most of their {core hydrogen-burning} lifetimes.
    Most metal lines are completely absent, including nitrogen. During their core helium-burning stage, lines switch to emission and even some metal lines (oxygen and carbon, but still almost no nitrogen) show up. Mass loss and clumping play a significant role in line-formation in later evolutionary phases, particularly during core helium-burning. 
	Most of our spectra are classified as an early O~type giant or supergiant, and we find Wolf--Rayet stars of type~WO in the core helium-burning phase.}
	{An extremely hot, early O~type star observed in a low-metallicity galaxy could be the outcome of chemically homogeneous evolution -- and therefore the progenitor of a long-duration gamma-ray burst or a type~Ic supernova.
	TWUIN~stars may play an important role in reionizing the Universe due to their being hot without showing prominent emission lines during the majority of their lifetimes.} 
	%We expect to obtain observational spectra of these stars by future campaigns aiming at the massive star content of low-metallicity star-forming galaxies e.g. \textit{Sextant~A} or \textit{I~Zwicky~18}.} 
	
	\keywords{stars: massive -- stars: winds, outflows -- stars: rotation -- galaxies: dwarf --  radiative transfer}
	
	\maketitle
	%
	%________________________________________________________________
	
	\pagebreak
	\section{Introduction}\label{sec:Intro}

Low-metallicity massive stars are essential building blocks of the Universe. Not only do these objects play a role in cosmology, contributing to the chemical evolution of the early Universe and the re-ionization history \citep[e.g.][]{Abel:2002,Yoshida:2007,Sobral:2015,Matthee:2018}. But also they may influence the structure of low-metallicity dwarf galaxies in the \textit{local} Universe \citep[e.g.][]{Tolstoy:2009,Annibali:2013,Weisz:2014}. Moreover, they may lead to spectacular explosive phenomena such as supernovae \citep[e.g.][]{Quimby:2011, Inserra:2013b, Lunnan:2013}, gamma-ray bursts \citep[e.g.][]{Levesque:2010, Modjaz:2011,Vergani:2015}, and possibly even gravitational wave emitting mergers \citep[e.g.][]{Abbott:2016a,Abbott:2017}. 
	The details of all these processes, however, are still weighted with many uncertainties. The reason for this is that low-metallicity ($<$~0.2~Z$_{\odot}$) massive stars have been rarely analyzed by quantitative spectroscopy as individual objects, since the instrumentation to obtain the required data quality has not been available for long.
	There are individual spectral {analyses} of massive stars only down to 0.1~{\Zsun}, such as for example one Wolf--Rayet (WR) star of type~WO in the galaxy IC~1613 \citep{Tramper:2013} and several hot stars in the galaxies IC~1613, WLM, and NGC~3109 \citep[e.g.][]{Tramper:2011,Herrero:2012,Tramper:2014, Garcia:2014, Bouret:2015,Camacho:2016}. Additionally, there are massive stars studied in the Small Magellanic Cloud (SMC) at Z$_{\mathrm{SMC}}$~$\sim$~0.2~{\Zsun}, including for example ten red supergiants \citep[][]{Davies:2015}, twelve WR stars \citep{Hainich:2015,Massey:2015,Shenar:2016}, and a few hundred O-type stars \citep{Lamb:2016}. 
	
	At metallicities below 0.1~\Zsun, however, there are so far no direct spectroscopic observations of individual massive stars. Although such stars might have been contributing to our Galaxy's chemical composition in the past \citep[specifically in globular clusters, see e.g.][]{Szecsi:2018,Szecsi:2018b}, they do not exist in our Galaxy anymore. Even if the second generation of stars in the early Universe was indeed composed of many massive and very massive stars \citep[e.g.][]{Choudhury:2007,Ma:2017}, our observing capacities are not sufficient to look that far for individual objects. 
	
	Even in local star-forming dwarf galaxies it is hard to resolve massive stars individually as they are embedded in dense and gaseous OB-associations \citep{Shirazi:2012, Kehrig:2013}. However, we may be able to find indirect traces of their existence such as the total amount of {ionizing photons} emitted by them, or the integrated emission lines of their WR stars \citep{Kehrig:2015, Szecsi:2015b, Szecsi:2015}. 
	Future observing campaigns may even provide us with a census of massive stars in metal-poor dwarf galaxies such as \textit{Sextant~A} \citep[$\sim$1/7~Z$_{\odot}$,][]{McConnachie:2012} or \textit{I~Zwicky~18} \citep[$\sim$1/40~Z$_{\odot}$,][]{Kehrig:2016}. 
    
   In this paper, we focus on a certain exotic type of low-metallicity massive stars: the fast-rotating, chemically-homogeneously evolving ones. \citet[][hereafter Paper\,I]{Szecsi:2015} called their core hydrogen-burning (CHB) phases TWUIN~stars; the term stands for \textit{Transparent Wind Ultraviolet INtense}. The reason why these stars were so named is that they were predicted to have weak, optically thin stellar winds, while being hot, and thus emitting most of their radiation in the UV band \citep[for more details see][]{Szecsi:2015b, Szecsi:2015, Szecsi:2016, Szecsi:2017short, Szecsi:2017long}. 
    TWUIN~stars have extensively been investigated from an evolutionary point of view, mainly as means to explain cosmic explosions and mergers. They were referred to as `stars with {chemically homogeneous evolution}' and `fast-rotating He-stars' by \citet{Yoon:2005} and \citet{Yoon:2006} who showed that they may be applied as single star progenitors of long-duration gamma-ray bursts and supernovae of type~Ib/c. They were referred to as `stars that evolve chemically homogeneously' by \citet{Brott:2011a} who presented such single star models with SMC metallicity. They were referred to as `the quasi-chemically homogeneous massive stars' by \citet{Cantiello:2007} who created such models to account for long-duration gamma-ray bursts, this time via binary interaction at Z$_{\mathrm{SMC}}$. They were referred to as `Wolf--Rayet stars in disguise' by \citet{deMink:2009b} who showed that such binaries may finally form a double black hole system. The latter hypothesis was further elaborated on by \citet{deMink:2016} and  \citet{Mandel:2016}, as well as \citet{Marchant:2016} and \citet{Marchant:2017} to provide progenitor channels to gravitational wave emission. In particular, \citet{Marchant:2016} found that chemically homogeneous stars at $\sim$0.02~Z$_{\odot}$ (indeed what we call TWUIN~stars here), when in a close binary, predict the highest rate of double black hole mergers compared to other metallicities. 
    
    All these authors above were mainly concerned either with the inner structure or with the final fate of these stars, but {rarely with their appearance}. In fact, sometimes theorists called them simply WR~stars \citep[e.g.][]{Cui:2018} as their surface composition and temperature, as predicted by the evolutionary models, are similar to those of observed WR~stars. However, to decide if they are in fact WR~stars from an observational point of view (i.e. if they show broad and
bright emission lines in the optical region), one should know their spectral appearance. A pioneer study in this direction was carried out recently by \citet{Hainich:2018}.
    
	This is the second paper of a series. In
	%\citet{Szecsi:2015},
	\citetalias{Szecsi:2015},
	stellar evolutionary computations of TWUIN~stars were presented (see their Sect.~6 and also~10.4) during the CHB phase, while some of these models were followed over the core helium-burning (CHeB) phase in \citet[][see Chapter~4 of the thesis]{Szecsi:2016}. In the current paper, we now simulate the atmospheres and spectra of chemically-homogeneously evolving stars of different masses, covering their whole evolution. We use the the Potsdam~Wolf-Rayet (\PoWR) stellar atmosphere code to compute the synthetic spectra. The initial metallicity of the evolutionary models based on which the synthetic spectra are created, is 0.02~Z$_{\odot}$. The choice of this particular metallicity value is motivated by the fact that binary models of this metallicity have been successfully applied in the context of double compact object progenitors \citep[e.g.][]{Marchant:2016} as well as other explosive phenomena \citep[see the review of][]{Szecsi:2017long}, and that such stars could potentially be found in some local dwarf galaxies \citep{Szecsi:2015b}. We explore expected observable characteristics of these stars, classify them accordingly and provide the spectral features that can be used to guide targeted observing campaigns. The predicted spectra will later be applied to create a synthetic population to be compared to observational properties of low-metallicity dwarf galaxies in a next part of this series. 

	This paper is organized as follows: 
	In Sect.~\ref{sec:stellarevolution} we give an overview of the stellar evolutionary model sequences used in this work. In Sect.~\ref{sec:stellaratmospere} we present the stellar atmosphere and wind models. In particular, stellar parameters and chemical composition are summarized in Sect.~\ref{sec:stellarparameters} while the wind properties are described in Sect.~\ref{sec:stellarparameters}. 
	In Sect.~\ref{sec:spectralmodels} we provide synthetic spectra of chemically-homogeneously evolving stars. The effects of mass loss and wind clumping on line formation are presented in Sect.~\ref{sec:masslossspectra} and Sect.~\ref{sec:clumping}, respectively. Classification of the model spectra are presented in Sect.~\ref{sec:spectralclasification}. In Sect.~\ref{sec:discussion}, we discus the validity of the models and suggest future research directions. Finally, a summary is given in Sect.~\ref{sec:conclusion}. All the calculated spectra are available in Appendix~\ref{sec:appendix}. 
	
	%__________________________________________________________________

	\section{Stellar evolutionary model sequences}\label{sec:stellarevolution}

	\begin{table*}\centering
		\caption{The main parameters of the 15 model stars in this study. An~*~marks those models that are undergoing CHeB (i.e. post-main-sequence evolution). log$\dot{M}$ refers to our nominal (`higher') mass-loss rate.
		{The columns C, N, and O show surface mass fractions of carbon, nitrogen, and oxygen, respectively.}
		We computed four synthetic spectra for each model in this table, corresponding to two different values of mass-loss rates (nominal and reduced) and clumping factors (D~$=$~1 and~10).}
		{\footnotesize 
			\begin{tabular}{lllllllllllllll}
				\hline\hline
				\rule[0mm]{0mm}{4.0mm}
				$\mini$ & label & $\log\Teff$ & $\log{\lstar}$
				& $\log\mdot$ & \YS & \YC & \ion{C}{} & \ion{O}{} & \ion{N}{} & \rstar & \mstar & $\log g$ &
				$v_\mathrm{rot}$ \\ %& $v_\mathrm{esc}$ \\
				$\hzav{\msun}$ & & $\hzav{\Kelvin}$ & $\hzav{\lsun}$ &
				$\hzav{\msunyr}$ & & & & & & $\hzav{\rsun}$ & $\hzav{\msun}$ &
				$\hzav{\cmss}$ & $\hzav{\kms}$ \\ %& $\hzav{\kms}$\\ 
				\hline
				%\rule[0mm]{0mm}{4.0mm}
				20 & 0.28 (T-1) & 4.58 & 4.68 & -8.48 & 0.28 & 0.34 & $5.47\cdot10^{-6}$ & $3.55\cdot10^{-5}$ & $9.12\cdot10^{-5}$ & 4.93 & 20.0 & 4.35 & 695 \\ %& 1244 \\
				20 & 0.50 (T-2) & 4.65 & 4.97 & -7.80 & 0.50 & 0.55 & $1.61\cdot10^{-6}$ & $2.88\cdot10^{-6}$ & $1.24\cdot10^{-4}$ & 5.01 & 20.0 & 4.34 & 675 \\ %& 1232 \\
				20 & 0.75 (T-3) & 4.74 & 5.29 & -6.89 & 0.75 & 0.78 & $2.13\cdot10^{-6}$ & $2.25\cdot10^{-6}$ & $1.24\cdot10^{-4}$ & 4.95 & 19.8 & 4.35 & 650 \\ %& 1237 \\
				20 & 0.98 (T-4) & 4.88 & 5.58 & -5.77 & 0.98 & 1.00 & $3.54\cdot10^{-6}$ & $1.59\cdot10^{-6}$ & $1.23\cdot10^{-4}$ & 3.58 & 19.2 & 4.61 & 702 \\ %& 1430 \\
				20 & CHeB (T-5) & 5.08 & 5.67 & -5.49 & 0.84 & 0.10* & $1.36\cdot10^{-1}$ & $2.13\cdot10^{-2}$ & $5.89\cdot10^{-3}$ & 1.55 & 16.8 & 5.28 & 994 \\ %& 2034 \\ \hline
				%\rule[0mm]{0mm}{4.0mm}
				59 & 0.28 (T-6) & 4.74 & 5.75 & -7.00 & 0.28 & 0.36 & $8.26\cdot10^{-6}$ & $4.00\cdot10^{-5}$ & $8.40\cdot10^{-5}$ & 8.14 & 58.9 & 4.39 & 421 \\ %& 1662 \\
				59 & 0.50 (T-7) & 4.79 & 5.94 & -6.70 & 0.50 & 0.57 & $2.27\cdot10^{-6}$ & $3.07\cdot10^{-6}$ & $1.23\cdot10^{-4}$ & 8.31 & 58.7 & 4.37 & 428 \\ %& 1642 \\
				59 & 0.75 (T-8) & 4.84 & 6.13 & -5.82 & 0.75 & 0.79 & $2.52\cdot10^{-6}$ & $1.94\cdot10^{-6}$ & $1.24\cdot10^{-4}$ & 8.08 & 58.3 & 4.39 & 422 \\ %& 1658 \\
				59 & 0.98 (T-9) & 4.92 & 6.29 & -4.92 & 0.98 & 1.00 & $3.94\cdot10^{-6}$ & $1.44\cdot10^{-6}$ & $1.23\cdot10^{-4}$ & 6.68 & 55.3 & 4.53 & 404 \\ %& 1778 \\
				59 & CHeB (T-10) & 5.14 & 6.34 & -4.70 & 0.68 & 0.10* & $2.41\cdot10^{-1}$ & $7.31\cdot10^{-2}$ & $3.65\cdot10^{-3}$ &  2.60 & 49.4 & 5.30 & 755 \\ %& 2694 \\ \hline
				%\rule[0mm]{0mm}{4.0mm}
				131 & 0.28 (T-11) & 4.76 & 6.29 & -6.17 & 0.28 & 0.30 & $3.34\cdot10^{-6}$ & $1.06\cdot10^{-5}$ & $1.15\cdot10^{-4}$ & 13.71 & 130.8 & 4.28 & 905 \\ %& 1908 \\
				131 & 0.50 (T-12) & 4.79 & 6.42 & -5.89 & 0.50 & 0.52 & $2.33\cdot10^{-6}$ & $2.06\cdot10^{-6}$ & $1.24\cdot10^{-4}$ & 14.26 & 129.9 & 4.24 & 925 \\ %& 1864 \\
				131 & 0.75 (T-13) & 4.84 & 6.57 & -4.96 & 0.75 & 0.76 & $2.71\cdot10^{-6}$ & $1.28\cdot10^{-6}$ & $1.24\cdot10^{-4}$ & 13.63 & 126.8 & 4.27 & 820 \\ %& 1884 \\
				131 & 0.98 (T-14) & 4.93 & 6.69 & -4.27 & 0.98 & 0.99 & $4.07\cdot10^{-6}$ & $1.39\cdot10^{-6}$ & $1.23\cdot10^{-4}$ &  10.18 & 112.5 & 4.47 & 520 \\ %& 2053 \\
				131 & CHeB (T-15) & 5.14 & 6.68 & -4.23 & 0.56 & 0.10* & $3.19\cdot10^{-1}$ & $1.23\cdot10^{-1}$& $3.79\cdot10^{-4}$ &  3.82 & 93.3 & 5.24 & 587 \\ %& 3050 \\ 
				\hline
				%\rule[0mm]{0mm}{4.0mm}
			\end{tabular}
		}
		\label{tab:list}
	\end{table*}

\begin{figure}\centering
		\includegraphics[angle=270,width=\columnwidth]{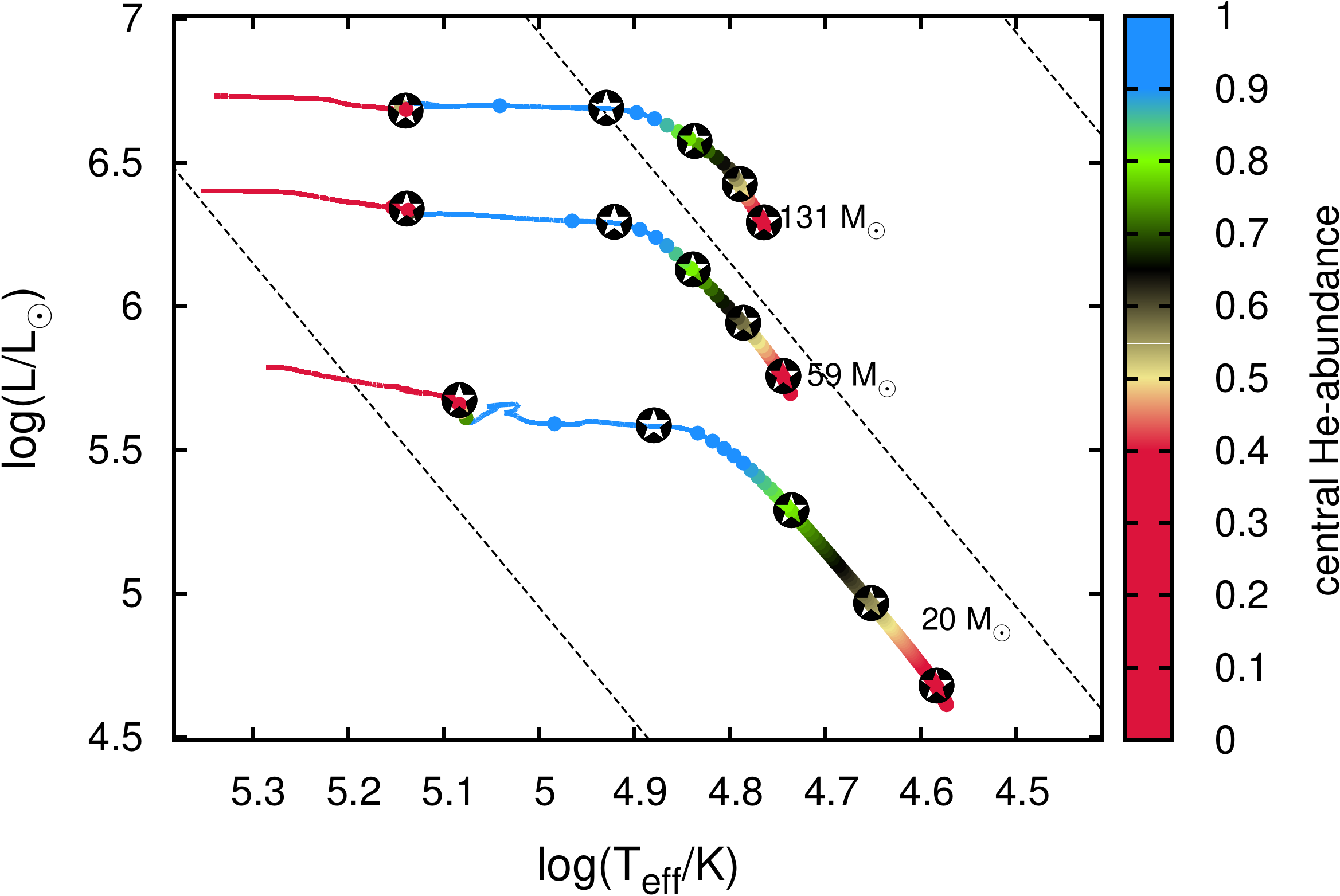}
		\caption{HR-diagram of our models (black symbols) and their corresponding evolutionary sequences. The sequences are taken from \citetalias{Szecsi:2015} and \citet{Szecsi:2016}. Initial masses are labelled, showing where the tracks start their evolution, proceeding towards the hot side of the diagram.
			Colors show the central helium mass fraction, while dots represent every 10$^5$~years of evolution. Dashed lines mark equiradial lines with 1, 10 and 100~R$_{\odot}$ from left to right, respectively.
			The black symbols represent those models that we computed synthetic spectra for. From the right to left, these black symbols correspond to evolutionary phases with surface helium mass fraction of 0.28, 0.5, 0.75 and 0.98; while the fifth one on the very left corresponds to the central helium mass fraction of 0.5, that is, the middle of the CHeB phase.
		}\label{fig:HRD}
	\end{figure}

	Single stellar evolutionary sequences of low-metallicity (Z$\sim$0.02~{\Zsun} or [\ion{Fe}{\!}/\ion{H}{\!}]~=~$-$1.7), fast-rotating massive stars were computed in  \citetalias{Szecsi:2015} during the CHB phase. The sequences were created using the Bonn Evolutionary Code (BEC). For the details of the code and the initial parameters of the computations, we refer to \citetalias{Szecsi:2015} and references therein. Since we are also interested in further hydrogen-free evolution, we rely on the work of \citet{Szecsi:2016} who continued the computation of these sequences during CHeB until helium exhaustion in the core. %In fact, the particular choice of the three initial masses is motivated by the computation of their CHeB phase in \citet{Szecsi:2016}.
	To represent different evolutionary stages with spectra, we use three chemically-homogeneously evolving sequences, namely those with initial masses {\mini} of $20\,\msun$, $59\,\msun$, and $131\,\msun$, and initial rotational velocities of $450\,\kms$, $300\,\kms$, and $600\,\kms$, respectively. These three tracks are shown in Fig.~\ref{fig:HRD}. 
	
	All three evolutionary sequences are computed assuming fast rotation initially, which is inherited to the 15 models that we compute spectra for. Their rotational velocities are in the range of $400-1000~\kms$ (see Table~\ref{tab:list}), which is still not close the critical rotational limit of these massive stars ($\sim$~0.4$-$0.6~$v_{\rm crit}$). Therefore, we do not expect these stars to form a decretion disc. Additionally, although these velocities may seem extremely high, a very similar evolution is found at lower rotational rates as well. For example, the model with $\mini=131\,\msun$ rotates with about $800-900\,\kms$ at the first part of its CHB lifetime, the model rotating only with
	$450\,\kms$ in this phase is evolving almost exactly the same way (cf. Fig.~4 in \citetalias{Szecsi:2015}). As discussed in Sect.~10.4 of \citetalias{Szecsi:2015}, it is expected that about 20\% of all massive stars at this metallicity evolve chemically homogeneously due to their fast rotation; indeed, observations down to Z$_{\mathrm{SMC}}$ suggest that stellar rotation increases with lower metallicity \citep{Mokiem:2006,Martayan:2007}. 
	
	For simulating the wind structure and spectra, we chose four models for each track, namely those with a surface helium mass fraction, {\YS}, of 0.28, 0.5, 0.75, and 0.98, as well as one model per track for the CHeB phase, as shown in Fig.~\ref{fig:HRD} (note the color coding in the figure showing the central helium mass fraction, {\YC}, of the model sequences, reflecting the evolutionary stage) and in Table~\ref{tab:list}. 
	
	%Brankica, is it possible to check how much a line (just one line is enough, say, for the model with 131Msun-0.98 the line C-IV-5808) would be different if rotational broadening is set instead of 520km/s to 400km/s? 
	
	\subsection{Mass loss applied in the evolutionary sequences}\label{sec:massloss}

	Mass-loss of massive stars may influence their evolution significantly even at this low metallicity (see \citetalias{Szecsi:2015}).
%\citep{Vink:2001,Szecsi:2015}    
The sequences were computed assuming a prescription for radiation-driven mass loss of hot O-type stars \citep{Vink:2000,Vink:2001} providing the mass-loss rate {\mdot} as a function of initial metal abundance {\Zini} (given in units of solar metallicity {\Zsun}) and further stellar parameters:
	\begin{multline}
		\log
		\frac{\mdot}{\msunyr} = -6.7 + 2.2 \log\zav{\lstar/10^5} - 1.3 \log\zav{\mstar/30}
		\\
		- 1.2 \log\zav{\frac{v_{\infty}/\vesc}{2.0}} + 0.9
		\log\zav{\Teff/40000}
		\\
		- 10.9 \hzav{\log\zav{\Teff/40000}}^2 + 0.85 \log(\Zini/\Zsun),
		\label{eq:vink}
	\end{multline}
	where {\mdot} is in units of $\msunyr$, stellar effective temperature {\Teff} is in units of Kelvin,  stellar mass {\mstar} and luminosity {\lstar} are in solar units; the ratio of the terminal velocity {\vinfty} and escape velocity {\vesc} is
	taken as
	$\vinfty/\vesc=2.6$ for 
	%our
	the evolutionary
	models as they all are above the bi-stability jump \citep{Lamers:1995,Vink:2000}. 
	%Dori, please correct me if I am not right.
	%
	This formula was applied when the {\YS} was lower than 0.55,
	which is true for every first two models of our three evolutionary sequences (i.e. T-1, T-2, T-6, T-7, T-11, and T-12 models, cf. Table~\ref{tab:list}).
	As the models are evolving chemically homogeneously, the surface
	abundances are very close to those in the core, $\YS \sim \YC$.
	
	A different prescription was assumed for phases when $\YS > 0.7$, which applies for WR stars, %as given below in Eq.~(\ref{eq:hamann}).
	\begin{equation}
	\log\frac{\mdot}{\msunyr}= 1.5\log \frac{\lstar}{\lsun} -2.85 X_{\rm S}
	-12.95 +0.85\log \frac{\Zini}{\Zsun},
	\label{eq:hamann}
	\end{equation}
	used for models T-3, T-4, T-8, T-9, T-13, and T-14.
	Here $X_{\rm S}$ is the surface hydrogen mass fraction.
	This expression follows from equation\,2 in \citet{Hamann:1995}, but has been reduced by a factor of 10 as suggested by \citet{Yoon:2006}. The reduction by 10 gives a mass-loss rate comparable to the commonly adopted one by \citet{Nugis:2000} \citep[see figure 1 in][]{Yoon:2015}. For the dependence on $X_{\rm S}$, see the steepness of the fit in figure~7 of \citet{Hamann:1995}. 
	
	%For phases between $0.55 < \YS < 0.7$ we interpolated linearly between the values given by the O-type mass-loss rate in Eq.~\eqref{eq:vink} and the WR-type mass-loss rate in Eq.~(\ref{eq:hamann}).
	%{\bf However, there are no such models in Table~\ref{tab:list}.}
	During the whole CHeB phase, the WR-type prescription of Eq.~\eqref{eq:hamann} was applied everywhere
	(models T-5, T-10, and T-15).
	
There are many uncertainties associated with this treatment of the wind mass loss.
For example, the prescription in Eq.~(\ref{eq:hamann}) includes a metallicity dependence of $\mdot\sim Z_{ini}^{0.85}$ following \citet{Vink:2001}. 
But in reality, the dependence may be weaker than that (i.e. real winds are stronger than assumed), as suggested by theoretical calculations for classic WR~stars in \citet{Vink:2005} and \citet{Eldridge:2006b}. Contrarily, observations of WN~stars carried out by \citet{Hainich:2015} found a stronger dependence (i.e. real winds are weaker than assumed). Thus it seems that the question of the metallicity dependence of WR~winds is yet to be settled.
%I marked this bold because this has been changed since we submitted.
%\hrule

Additionally, WN~stars and WC~stars may well be different from each other when it comes to wind mass loss; and both quite different from the CHB phase of our chemically-homogeneously evolving models (when they are TWUIN~stars). Still, the reason of \citetalias{Szecsi:2015} for using a mass-loss rate prescription based on observations of WR~stars to simulate TWUIN~stellar evolution, was that in terms of surface composition and temperature, WR~stars are the closest objects to TWUIN~stars. We provide suggestions for future research directions to establish the wind properties of TWUIN~stars (both observationally and theoretically) in Sects.~\ref{sec:theory} and~\ref{sec:observations}.

%There might also be quite a difference between WN stars and WC stars, but in this paper we only deal with something that would at most be comparable to WN stars, so referring to the scaling obtained by \citet{Hainich:2015} is well motivated.

To account for all these uncertainties,
we created two versions for every model. One has a \textit{nominal} mass-loss rate as implemented in the evolutionary models, i.e. according to Eqs.~(\ref{eq:vink}) and~(\ref{eq:hamann}). Another one has a \textit{reduced} value which is a factor of hundred lower than the nominal value. Choosing a factor of hundred is motivated by the work of \citet{Hainich:2015} who found steeper metallicity-dependence of WR winds; i.e. using the mass-loss prescription given by equation~11 of \citet{Hainich:2015}, we obtained mass-loss rates close to our reduced values, see Table~\ref{tab:massloss}. The nominal value is henceforth usually referred to as `higher', which means in context of our study that it is the higher one of the two. By testing these two, rather extreme values, we account for uncertainties concerning the mass-loss predictions of these stars.

	\begin{table}[t!]
    \centering
		\caption{Mass-loss rate values applied in the synthetic spectra computations, compared to those that the work of \citet{Hainich:2015} would predict for the same stars. $\log{{\mdot}_\mathrm{r}}$ means the reduced mass-loss rate, and $\log{{\mdot}_\mathrm{h}}$ the `higher', i.e. the nominal one as applied in the evolutionary sequences in Paper\,1 (that is, computed using our Eq.~(\ref{eq:vink}) or~(\ref{eq:hamann})).}
		%\vspace{20pt}
		\begin{tabular}{llccc}\hline\hline\small
		\rule[0mm]{0mm}{4.5mm}
		  {\mini} & label & $\log{\mdot_\mathrm{h}}$ & $\log{{\mdot}_\mathrm{r}}$ & $\log{\mdot_\mathrm{Hainich}}$ \\
		  & & $\hzav{\msunyr}$ & $\hzav{\msunyr}$ & $\hzav{\msunyr}$ \\
			\hline
			20 & T-1 (0.28)  & -8.48    & -10.48   & -10.77   \\
20 & T-2 (0.50)  & -7.80    & -9.80    & -10.39   \\
20 & T-3 (0.75)  & -6.89    & -8.89    & -8.68    \\
20 & T-4 (0.98)  & -5.77    & -7.77    & -7.99    \\
20 & T-5 (pMS)   & -5.50    & -7.50    & -8.17    \\
\hline
59 & T-5 (0.28)  & -7.00    & -9.00    & -9.29    \\
59 & T-6 (0.50)  & -6.70    & -8.70    & -8.29    \\
59 & T-7 (0.75)  & -5.82    & -7.82    & -7.52    \\
59 & T-8 (0.98)  & -4.92    & -6.92    & -7.00    \\
59 & T-10 (pMS)  & -4.70    & -6.70    & -7.50    \\
\hline
131 & T-11 (0.28)& -6.17    & -8.17    & -8.53    \\
131 & T-12 (0.5) & -5.89    & -7.89    & -8.04    \\
131 & T-13 (0.75)& -4.96    & -6.96    & -9.10    \\
131 & T-14 (0.98)& -4.27    & -6.27    & -7.80    \\
131 & T-15 (pMS) & -4.23    & -6.23    & -7.27    \\
\hline
		\end{tabular}
		%\\ $\star$ means Fe2-10
		\label{tab:massloss}
	\end{table}
	
	%______________________________________________________________
	\section{Stellar atmosphere and wind models}\label{sec:stellaratmospere}
	
	To calculate the synthetic spectra and to obtain the stratification of wind parameters, a proper modelling of the static and expanding atmosphere is required. We calculated  stellar spectra by means of the state-of-the-art Potsdam~Wolf-Rayet (\PoWR) atmosphere code. As the {\PoWR} code treats both quasi-static (i.e photospheric) and expanding layers (i.e. wind) of the stellar atmosphere consistently, it is applicable to most types of hot stars.
	
	The {\PoWR} code solves the non-LTE radiative transfer in a spherically expanding atmosphere with a stationary mass outflow. A consistent solution for the radiation field and the population numbers is obtained iteratively by
	solving the equations of statistical equilibrium and radiative transfer in the co-moving frame \citep{Mihalas:1978, Hubeny:2014}. After an atmosphere model is converged, the synthetic spectrum is calculated via a formal integration along emerging rays.
	
	To ensure energy conservation in the expanding atmosphere, the
	temperature stratification is updated iteratively using the electron
	thermal balance method \citep{Kubat:1999} and a generalized form of the so-called Uns{\"o}ld-Lucy method assuming radiative equilibrium \citep{Hamann:2003}. In the comoving frame calculations during the non-LTE iteration, the line profiles are assumed
	to be Gaussians with a constant Doppler broadening velocity {\vdop},
	which accounts for broadening due to thermal and microturbulent velocities. In this work we use $\vdop=100~\kms$. All spectra correspond to being seen edge-on, i.e. lines are fully broadened by rotation.
	
	After the model iteration converged and all population numbers have been established, the emergent spectrum is finally calculated in the observer's frame, using a refined set of atomic data (e.g. with multiplet splitting) and accounting in detail for thermal, microturbulent and pressure broadening of the lines.
	
	Detailed information on the assumptions and numerical methods used in the code can be found in \citet{Grafener:2002}, \citet{Hamann:2003, Hamann:2004}, and \citet{Sander:2015}.
	
	\subsection{Stellar parameters and chemical composition}\label{sec:stellarparameters}
	
	Fundamental stellar parameters required as input for {\PoWR} model atmosphere calculations are the stellar temperature $T_\ast$, the stellar mass \mstar, and the stellar luminosity {\lstar}. These are adopted from the stellar evolutionary model sequences (see Table~\ref{tab:list}), assuming that the hydrostatic surface temperature {\Teff} of the BEC evolutionary models coincides with {\tstar}.
	With given {\lstar} and $T_\ast$,
	the stellar radius {\rstar} is calculated via Stefan-Boltzmann’s law
	\begin{equation}
	\lstar=4\pi\sigma_{\mathrm{SB}}\rstar^2{\tstar}^{4},
	\label{eq:SBlaw}
	\end{equation}
	where $\sigma_{\mathrm{SB}}$ is the Stefan-Boltzmann constant. In the {\PoWR} code the  temperature {\tstar} is an effective temperature at the radius {\rstar}, which is defined at the Rosseland continuum optical depth $\tau_{\mathrm{max}}=20$. The outer atmosphere (i.e. wind) boundary is set to $1000~\rstar$ with the exception of the models for $M_\text{ini} = 20$~\msun where $100~\rstar$ is already sufficient. Further details about the method of model atmosphere calculations can be found in \citet{Sander:2015}.
	
	%
	% (Dori) I have a question here. We know that WR stars have their photosphere in the wind, not on the hydrostatic surface. So, the effective temperature of WR stars are lower than their temperature on the surface. This is not important for stars with only absorption lines, but may be important for stars with emission lines? I mean, their position on the HR diagram may change due to their wind being optically thick, right? Should we not indicate this on Fig.1?
	%
	
	Detailed model atoms of all relevant elements are taken into account. Line blanketing is considered with the iron-group elements treated in the super-level approach, accounting not only for \ion{Fe}{\!}, but also Sc, Ti, V, Cr, Mn, Co,
	and Ni \citep[see][for details]{Grafener:2002}. The abundances of \ion{H}{\!}, \ion{He}{\!}, \ion{C}{\!}, \ion{N}{\!}, \ion{O}{\!}, \ion{Ne}{\!}, \ion{Mg}{\!}, \ion{Al}{\!}, \ion{Si}{\!}, and \ion{Fe}{\!} are adopted from the stellar evolutionary model sequences. The additional elements such as \ion{P}{\!}, \ion{S}{\!}, \ion{Cl}{\!}, \ion{Ar}{\!},
	\ion{K}{\!}, and \ion{Cl}{\!}, which are not considered in the stellar evolutionary models but used in the {\PoWR} model atmosphere calculations, are also considered in a minimal level approach to account for their potential contributions to the wind driving. The additional elements have abundances of $\Zsun/50$. For iron group elements we consider the ionization stages from \ion{\!}{i} up to \ion{\!}{xvii} to ensure that all sources which significantly contribute to opacity are taken into account. Higher ionization stages of \ion{Fe}{\!} are important especially for the CHeB stages of the considered stars.
	
	\begin{figure*}[htp]\centering
		\includegraphics[width=0.9\textwidth]{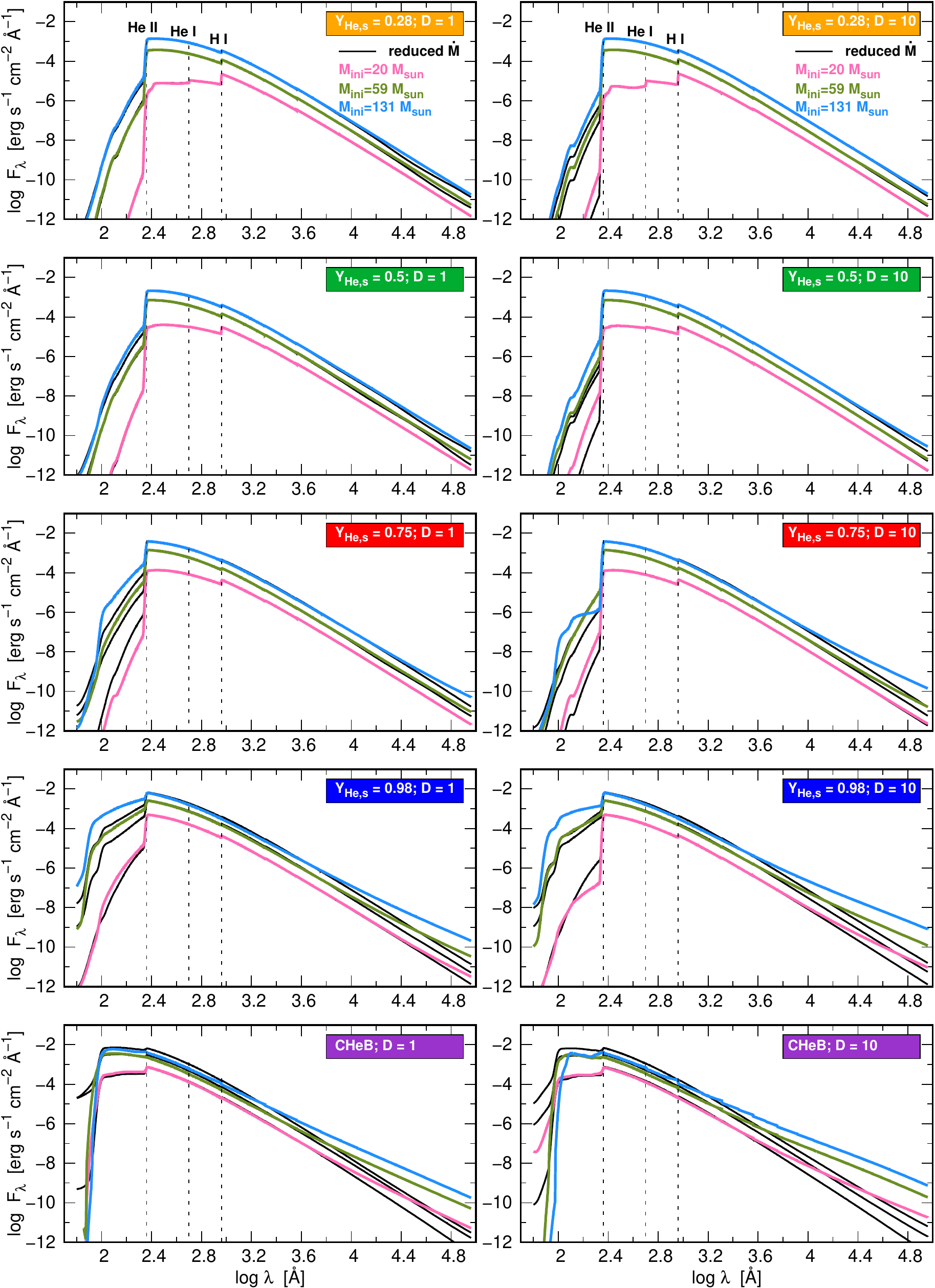}
		\caption{Spectral energy distribution (continuum) of chemically-homogeneously evolving stars
			%(with different  $\mini$ colored lines as denoted in the uppermost panels) 
			in different evolutionary stages as marked in the colored boxes in each panel.
			The left panels provide the continuum SED of the models calculated with a smooth ($D=1$) wind, while the right panels depict the same for the clumped ($D=10$) wind assumption. The colored lines correspond to the models with specific {\mini} (denoted in the uppermost panels) and calculated assuming the same (nominal) mass-loss rate as given in Table~\ref{tab:list}. For each colored line, there is also a black line representing the SED of the model for the same star in the same evolutionary stages with the same clumping factor $D$, but assuming a mass-loss rate hundred times lower. For the better visibility of differences between the SEDs in the CHeB phase see Appendix Fig.~\ref{fig:SED-D1-D10-evol-cheb}%{\it The SED curve which corresponds to the blue line in the lowermost right panel is not flux-consistent}.
			%\textbf{\color{violet}Mention which model is not flux-consistent.}
			%The mass-loss rates are taken from the stellar evolutionary calculations and wind is assumed to be smooth ($D=1$, left panels) or clumped ($D=10$, right panels). Black lines correspond to models of the same stars in the same evolutionary stages but calculated assuming mass-loss rates by two orders of magnitude lower.
		}
		\label{fig:sed-evol}
	\end{figure*}
	
	\subsection{Wind properties}\label{sec:windproperties}
	
	As we are dealing with objects which were predicted only theoretically and have never been observed, there exist no observational constraints on their wind properties yet.
	Within the frame of model consistency there is, therefore, some freedom in adopting atmospheric and wind parameters.
	
	\paragraph{Mass-loss rates}
	
	With specified $\mdot$ in the {\PoWR} code, the density stratification $\rho(r)$ in the wind is calculated via the continuity equation given as
	\begin{equation}
	\mdot=4\pi r^2\,\vel(r)\,\rho(r).
	\end{equation}
	
	To be consistent with stellar evolutionary models that provide the basis for our spectral models, we decided to apply the same mass-loss rate values as in those models.
	Note that these values were \textit{assumed} in the evolutionary
	models based on prescribed recipes (see Sect.~\ref{sec:massloss}), and \textit{not predicted} by those models. Therefore, to test the effect of mass loss on the emergent spectra, we supplement our work by another set of
	models: one calculated with a mass-loss rate which is hundred times lower than in the original set (see Table~\ref{tab:massloss} and Sect.~\ref{sec:massloss}).
	This enables us to roughly estimate uncertainties of our emergent radiation
	prediction due to uncertainties in the choice of mass-loss rates.
	
	\paragraph{Velocity}
	The adopted velocity field in the {\PoWR} models consists of two parts. A hydrostatic part where gravity is balanced by gas and radiation pressure, and a wind part where the outward pressure exceeds gravity and therefore the matter is accelerated. To properly account for the velocity field in the inner part of the wind, the quasi-hydrostatic part of the atmosphere is calculated self-consistently to fulfill the hydrostatic equation. Computing hydrodynamically consistent stellar atmosphere models this way is a new approach, recently implemented in the {\PoWR} code \citep[see][]{Sander:2015}.
	In the wind domain (i.e. the supersonic part) the velocity field is prescribed by the so-called $\beta$-law \citep[see, e.g.,][]{Lamers:1999} as 
	\begin{equation}
	\vel(r) = \vinf \zav{1-\frac{\rstar}{r}}^{\beta}
	\label{betavel}
	\end{equation}
	where $\vinf$ is the wind terminal velocity and $\beta$ a parameter describing the steepness of the velocity law.
	
	Since there exist no predictions (neither theoretically based nor observationally implied)
	about the velocity field of TWUIN~stars, we adopt only schematic parameters. 
	For $\beta$ we assume values of $0.8$ or $1.0$. This choice is motivated by the fact that the $\beta$ parameter's typical value for massive stars ranges between 0.6 and 2.0 \citep[see e.g.][]{Puls:2008}.
	As for the terminal wind velocity $\vinf$, we assume the same value for all models, that is, $\vinf=1000\,\kms$.
	This is a reasonable estimate, since in the simplified relation between terminal and escape velocities ($\vinf/\vesc=2.6$) used in mass-loss rate prescriptions the ratio $\vinf/\vesc$ lowers significantly when accounting for a rapid stellar rotation \citep{Friend:Abbott:1986}.
	In Sect.~\ref{sec:validity} we discuss possible ways to improve the assumptions about ${\vinf}$ in the future.

	\paragraph{Clumping} 
	As clumping is another wind property which influences the emergent spectra, we also calculated an additional set of models assuming the presence of clumping in the 
	wind. This enables us to estimate the influence of clumping on our prediction of the
	emergent radiation.
	
	Wind inhomogeneities are treated in the microclumping 
	approximation \citep[see][]{Hamann:1998}, which means that all clumps are assumed to be
	optically thin. The density in clumps is enhanced by a clumping factor 
	$D=1/f_\text{V}$, where $f_\text{V}$ is a fraction of volume occupied by clumps (i.e. volume filling factor). 
	The inter-clump medium is assumed to be void. For models in which clumping was 
	assumed, we also allow the clumping factor to depend on radius. We implement clumping stratification with 
   \begin{equation}
   f_{\text{V}}(r)=f_{\text{V},\infty}+(1-f_{\text{V},\infty})\exp{\left(-\frac{\tau_{\mathrm{cl}}}{\tau_{\mathrm{Ross}}(r)}\right)},
   \end{equation}
   where $f_{\text{V},\infty}=D_\infty^{-1}$, $D_\infty$ denotes the maximum clumping value, and $\tau_{\mathrm{cl}}$ is a free parameter denoting a characteristic Rosseland optical depth for the clumping ``onset'' \citep[for more details see][]{Sander:2017}. In all models with a depth-dependent clumping stratification we use $\tau_{\mathrm{cl}} = 2/3$.
    %, thereby starting to deviate from a homogeneous (i.e. smooth) wind with $D=1$ at about the sonic point (5~\kms) and quickly reaching $D=10$ at $\vel=40~\kms$.

	%\begin{figure*}[htp]\centering
	%\includegraphics[width=0.93\textwidth]{SED-D1-D10-evol.pdf}
	%	\caption{Spectral energy distribution (continuum) of the models with $\mini=20~\msun$, 59~$\msun$, and 131~$\msun$ (with colors given in the title of the figure) in different evolutionary stages defined with the value $\YS$ as it marked in the panels. The mass-loss rates are taken from the stellar evolutionary calculations and wind is assumed to be smooth (e.i. $D=1$). The black lines correspond to the spectral energy distribution (continuum) of the same stars in the same evolutionary stages but calculating assuming the two orders of the magnitude lower mass-loss rates.}
	%	\label{fig:sed-evol}
	%\end{figure*}
	
	%______________________________________________________________
	
	\section{Spectral models}\label{sec:spectralmodels}
	
	To explore the spectral appearance of chemically-homogeneously evolving stars, we computed four sets of atmosphere models with three different {\mini} ($20$, $59$, and $131\,\msun$) and for five different evolutionary stages defined by $\YS$ (0.25, 0.5, 0.75, 0.98, and CHeB). The models of the CHeB evolutionary phase have no hydrogen and their $\YS$ abundances are given in Table~\ref{tab:list}. The four sets of models consist of two sets with different values of a mass-loss rate and two sets with different values of a clumping factor. In total we created 60 models. 
	
	For the calculation of line profiles, we took into account line broadening for all lines, accounting for radiation damping, pressure broadening as well as rotational broadening. For the latter, we used the same value of the rotational velocity $\vel_{\mathrm{rot}}$ as in stellar evolutionary models. The influence of rotation on line formation is usually accounted for by performing a flux-convolution with a rotation profile. However, this may not be valid in the case of expanding atmospheres. Therefore, we used an option in the {\tt PoWR} code  which accounts for rotation with a 3D integration scheme of the formal integral assuming the co-rotation radius same as the radius of the star \citep[for more details see][]{Shenar:2014}. The mass-loss rates and rotational velocities used in the calculations are given in Table~\ref{tab:list}. 
	
	The continuum spectral energy distributions (SEDs) of all models are shown in Fig.~\ref{fig:sed-evol}. 
	The maximum emission is found in the far and extreme ultra-violet (UV) region. 
	With increasing {\mini}, also the luminosity and thus the resulting flux increases. With the exception of the $20\,\msun$ model at the first evolutionary stage, the flux maximum is always close to the \ion{He}{ii} ionization edge. %The shift of the SED curves towards higher fluxes is the smallest in the CHeB phase.
%\textbf{\textsl{\color{violet}Maybe here we should write about model TWUIN-59-0.98}}. 
%The A not so prominent shift is also seen at larger wavelengths in the visible and in the infrared spectral regions (see colored lines in the left panels of Fig.~\ref{fig:sed-evol}). 
	
	The SEDs also reveal that for all three mass branches the amount of emitted far and extreme UV ionizing radiation increases more and more during the evolution of the stars. This is a direct consequence of the  chemically homogeneous evolution where $\Teff$ is most of the time increasing monotonically.
%{\it In the optical regime however, there is a wavelength range where this trend is reversed, meaning that a more evolved star is less bright in this range, even though the total luminosity increases. This effect is not exclusive to TWUIN~stars and e.g. has to be taken into account when analyzing WR+O binary systems where the WR component has a higher luminosity but a lower visual magnitude than its component \citep[see, e.g.,][]{Shenar:2016}}. \textbf{\color{violet}Why is this italic font?} -- Okay, we figured it out. It has nothing to do with the physics of the wind. It's because the 59-0.98 model is slighly more evolved.
	
	Lowering the mass-loss rates has no significant influence on the emitted radiation during most of the CHB phases. Only small differences in the emitted fluxes can be seen at wavelengths shorter than 227\,$\angstrom$ and longer than 10\,000\,$\angstrom$ (see differences between colored and black lines in the left panels of Fig.~\ref{fig:sed-evol}). The same conclusion can be drawn for the clumped wind models with $D=10$ (see differences between colored and black lines in the right panels of Fig.~\ref{fig:sed-evol}). Those differences are higher and more visible in the evolutionary stages shortly before the end of CHB phase and in the CHeB phase (see differences between colored and black lines in the left and right panels with blue and purple boxes in Fig.~\ref{fig:sed-evol}). 

Differences in the emitted fluxes between models calculated for smooth and clumped wind are very small and present mostly at the wavelengths shorter than 227\,$\angstrom$, independently of the adopted {\mdot}. Small differences between SEDs are also found at wavelengths longer than 10\,000\,$\angstrom$ for models in the later stages assuming higher {\mdot} (see differences between black and colored lines in the left panels with higher {\mdot} and right panels with reduced {\mdot} in Appendix Fig.~\ref{fig:SED-D1-D10-evol-clump}).
	
	The SEDs reveal that the radiation with frequencies higher than the \ion{H}{i}, \ion{He}{i}, and \ion{He}{ii} ionization limits increase both with the initial mass and during the evolution of the stars. More massive and more evolved stars emit more ionizing flux. The consequences of ionizing fluxes of chemically-homogeneously evolving stars and their application will be discussed in a subsequent paper (Sz\'ecsi et al., in prep).
	
	\begin{figure*}[htp]\centering
		\includegraphics[width=1.28\textwidth, angle =90]{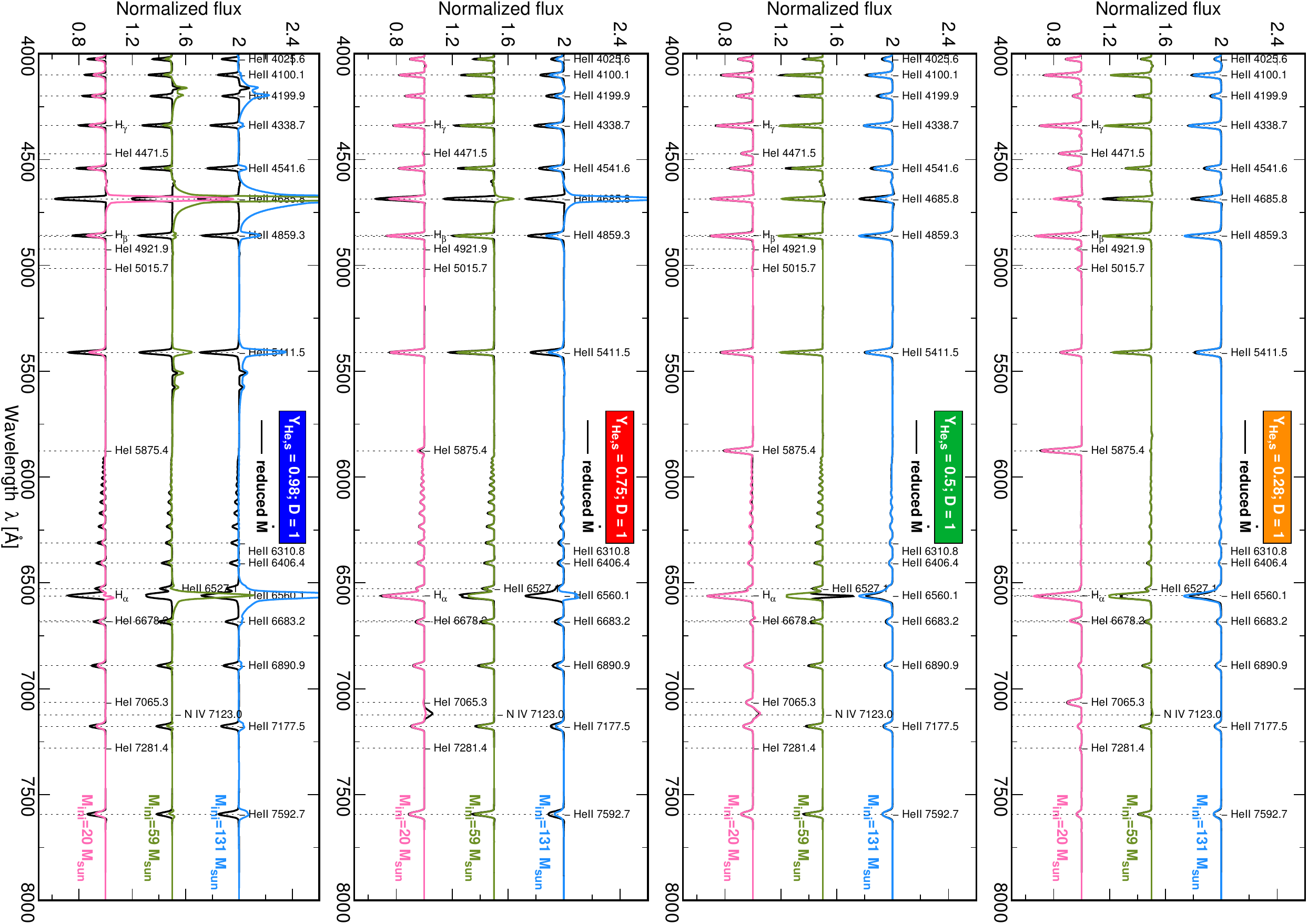}
		\caption{{\PoWR} spectra in the optical region of TWUIN~stars with different $\mini$ (see labels on the right sides of the panels) and in different CHB evolutionary phases marked by the value $\YS$ in the colored boxes. The colored lines correspond to the models calculated with mass-loss rates as given in Table~\ref{tab:list}. The black lines correspond to the models of the same stars in the same evolutionary stages but calculated with mass-loss rates hundred times lower (i.e. reduced {\mdot}). In all cases, the spectra correspond to smooth ($D=1$) wind models.
        }
		\label{fig:MS-opt-D1}
	\end{figure*}
	
	\begin{figure*}[htp]\centering
		\includegraphics[width=1.\textwidth, angle =90]{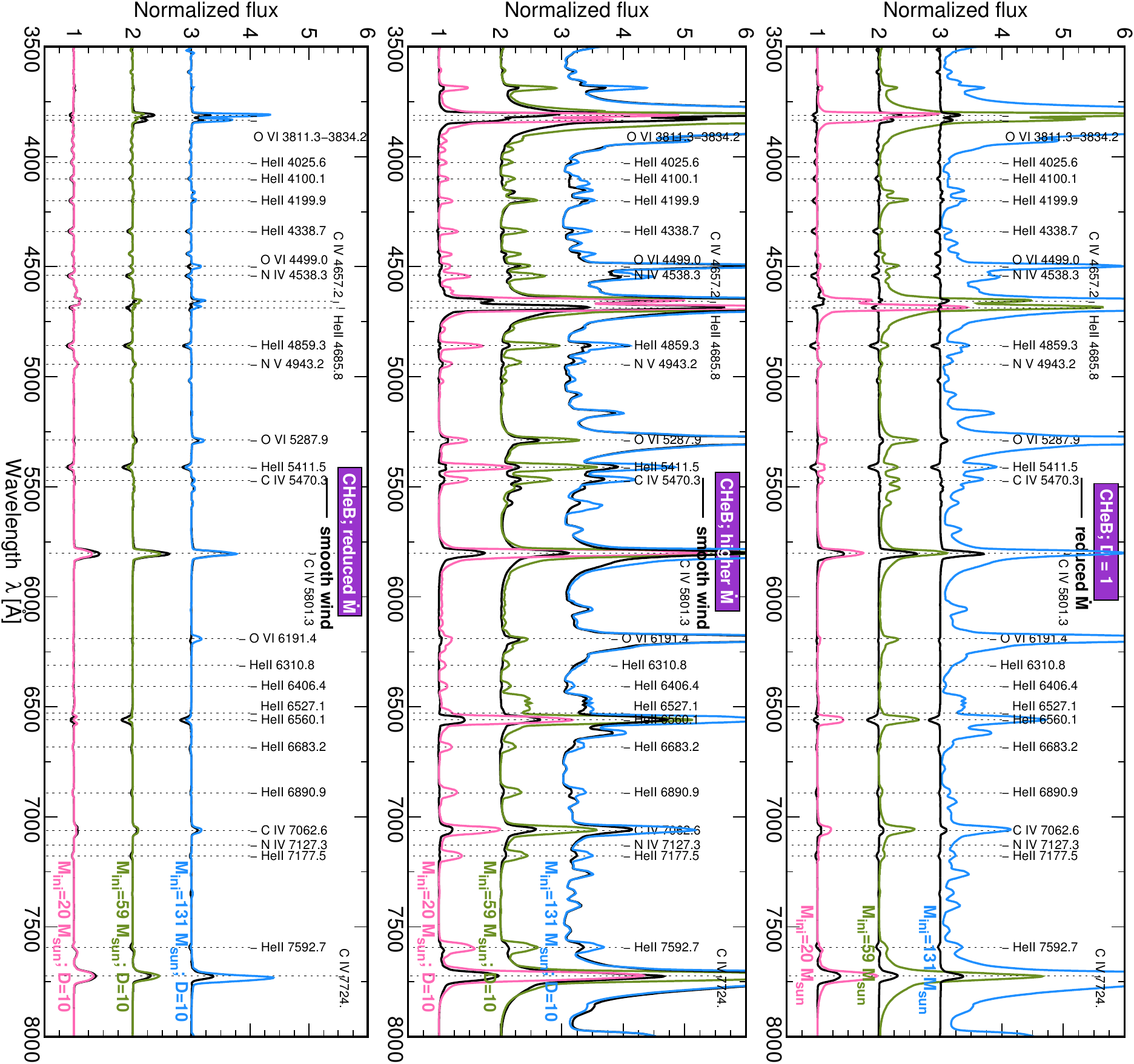}
		\caption{Uppermost panel: same as Fig.~\ref{fig:MS-opt-D1}  but for the CHeB evolutionary phase with $\YS$ as given in Table~\ref{tab:list}. Middle and lowermost panels: same as the uppermost panel but for clumped wind (i.e. $D=10$)  with nominal (i.e. higher) {\mdot} (the middle panel) and reduced {\mdot} (the lowermost panel); black lines correspond to the model with smooth wind assumption.}
		\label{fig:pMS-opt-D1-D10-r}
	\end{figure*}
	
	\begin{figure*}[htp]\centering
		\includegraphics[width=0.95\textwidth]{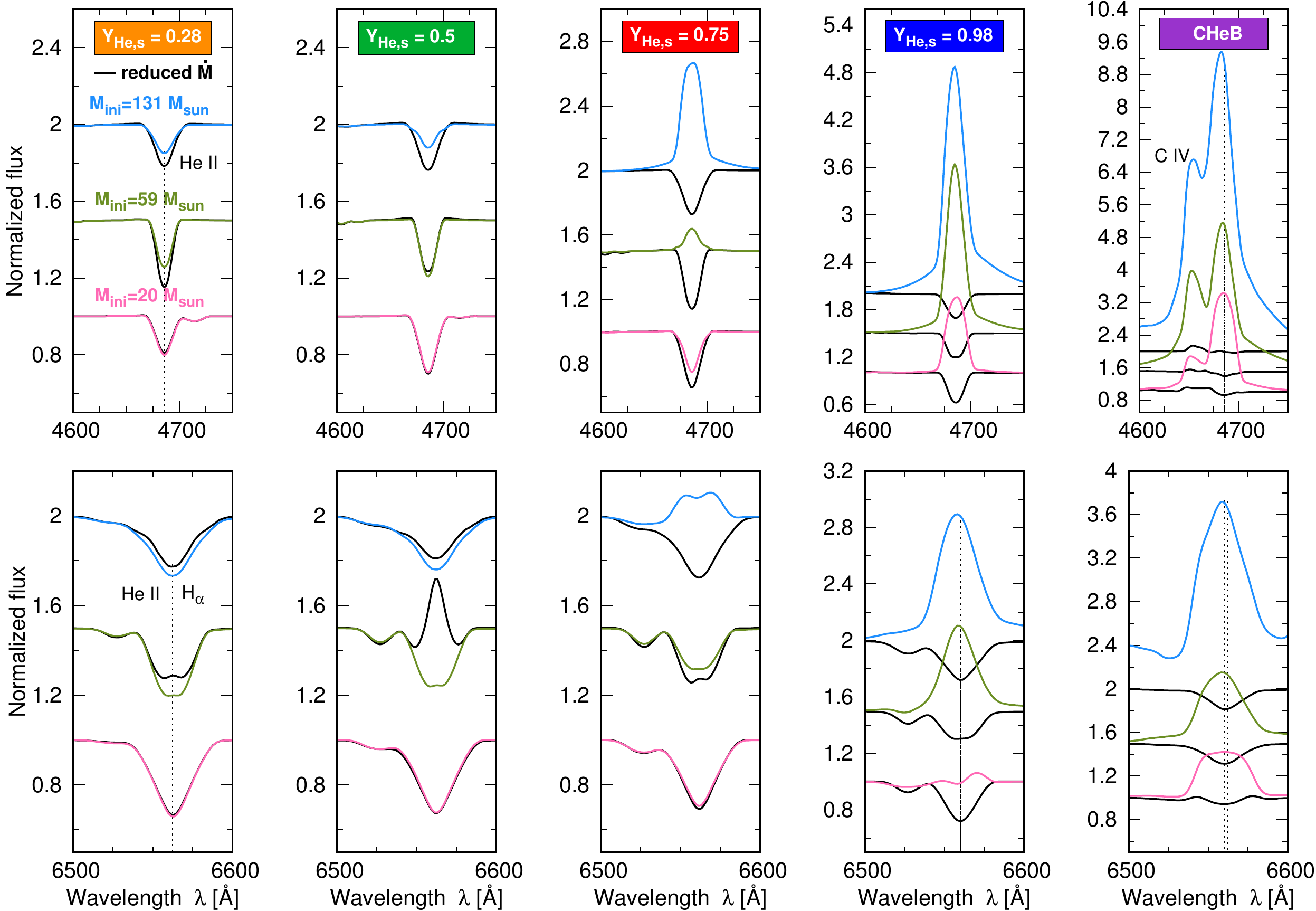}
		\caption{Same as Fig.~\ref{fig:MS-opt-D1} but zooming in on the \ion{He}{ii} $\lambda4686\,\angstrom$ line (upper panels) and {\Halpha} line blended with \ion{He}{ii} $\lambda6560\,\angstrom$ line (lower panels). In the CHeB evolutionary phase \ion{C}{iv} $\lambda4657\,\angstrom$ line appear.} 
		\label{fig:stronglines}
	\end{figure*}

	\subsection{Description of spectral features}\label{sec:spectra}
	
	To discuss the detailed spectral features, we analyze the normalized spectra. The optical range is depicted in Figs.~\ref{fig:MS-opt-D1} and \ref{fig:pMS-opt-D1-D10-r} while the spectra in the UV and infrared (IR) regions of each model are plotted in Appendix Figs.~\ref{fig:MS-uv-D1}~--~\ref{fig:pMS-ir-D1-D10-r}.
	
	The spectra calculated with mass-loss rates taken from stellar evolution calculations and assuming a smooth wind in earlier evolutionary phases with CHB, show most lines in absorption (see colored lines in Fig.\,\ref{fig:MS-opt-D1}). These lines turn into emission in the CHeB phase, during which these stars have no hydrogen in the atmosphere (see colored lines in the uppermost panel in Fig.~\ref{fig:pMS-opt-D1-D10-r}). This trend is also visible in the UV and IR spectra (see also colored lines in Appendix Figs.~\ref{fig:MS-uv-D1} and \ref{fig:MS-ir-D1} and in the uppermost panels in Figs.~\ref{fig:pMS-uv-D1-D10-r} and \ref{fig:pMS-ir-D1-D10-r}). The spectra of these stars do not typically show any P Cygni line profiles. This is somehow surprising, but this is most likely related to low Z of these stars.
	
	Synthetic spectra of models with $\YS=0.25$ and $\YS=0.5$, independently of their initial mass, show almost exclusively absorption lines in most of the spectral regions, except the emission line \ion{N}{iv}~$\lambda7123\,\angstrom$ in the optical range (see the pink line in the panel with green box in Fig.~\ref{fig:MS-opt-D1}); and very weak blending lines \ion{He}{ii}~$\lambda1.86\,\microm$ and \ion{He}{i}~$\lambda1.88\,\microm$ in the IR range (see purple and green lines in the panels with orange and green boxes in Appendix Fig.~\ref{fig:MS-ir-D1}). 
	
	When the stars reach the evolutionary stage with $\YS=0.75$, some additional emission lines appear in the spectra of the higher-mass models ({\mini}~=~$59\,\msun$ and $131\,\msun$). For instance, 
	helium emission line \ion{He}{ii}~$\lambda4686\,\angstrom$ and hydrogen emission line \Halpha~$\lambda6563\,\angstrom$ can be found in the optical spectra (see the green and blue lines in the panel with red box in Fig.~\ref{fig:MS-opt-D1}). In the UV spectral region, helium emission line \ion{He}{ii}~$\lambda$$1641\,\angstrom$ and nitrogen line \ion{N}{v}$~\lambda$$1239\,\angstrom$ can be found (see colored line in the panel with red box in Appendix Fig.~\ref{fig:MS-uv-D1}). In the IR part of the spectra, additional \ion{He}{ii} emission lines (for instance \ion{He}{ii} $1.01\,\microm$ and \ion{He}{ii} $1.16\,\microm$) can be found only in the spectra of the highest-mass model (i.e. {\mini}~=~$131\,\msun$, see the blue line in panel with red box in Appendix Fig.~\ref{fig:MS-ir-D1}). 
	The model with $\mini~=~20\,\msun$ does not show any sign of emission lines in this evolutionary phase. Even the emission line \ion{N}{iv}~$\lambda7123\,\angstrom$ disappears.
	
	At the evolutionary stage with $\YS=0.98$, i.e. shortly before the end of CHB, synthetic spectra of the higher-mass models with $\mini=59\,\msun$ and $131\,\msun$ show more intense emission lines. In addition to the emission lines they had in the previous evolutionary phases, more \ion{He}{ii} lines in all spectral regions are now in emission
	%, such as \ion{He}{ii}~$\lambda4542.73\,\angstrom$, \ion{He}{ii}~$\lambda4860.54\,\angstrom$, \ion{He}{ii}~$\lambda5411.53\,\angstrom$, and \ion{He}{ii}$~\lambda8236.51\,\angstrom$ 
	(see colored lines in the panels with blue boxes in Fig.~\ref{fig:MS-opt-D1}, and in Appendix Figs.~\ref{fig:MS-uv-D1} and \ref{fig:MS-ir-D1}). In the UV spectral region, a hydrogen line L$\alpha$~$\lambda1216~\angstrom$ appears in emission. This line would be probably masked by interstellar absorption when observed in the local Universe; but at high-redshift, provided that a sufficiently massive population of chemically-homogeneously evolving stars are present, it may indeed be identifiable in the host galaxy's spectra.
	
	In addition, other \ion{He}{ii} lines as well as metal lines of \ion{C}{iv} and \ion{O}{vi} appear (see colored lines in the panel with blue box in Appendix Fig.~\ref{fig:MS-uv-D1}). As for \ion{N}{v} lines, only \li{N}{V}{4606} shows up in absorption, every other nitrogen line is completely absent.
	In the IR spectral regions, more \ion{He}{ii} lines are now seen in emission
	%, including for instance now also \ion{He}{ii}~$\lambda=1.01\,\microm$ and \ion{He}{ii}~$\lambda=1.16\,\microm$ 
	(see colored lines in the panel with blue box in Appendix Fig.~\ref{fig:MS-ir-D1}). Lines which were in emission in the previous evolutionary phase now became much stronger. The spectra with $\mini~=~20\,\msun$ also show these emission lines at this evolutionary stage.  
	
	The strongest emission line in the optical spectra up to this evolutionary stage is the \ion{He}{ii}~$\lambda4686\,\angstrom$ line. The flux in the line center corresponds up to about 2 times that of the continuum (see zoom of the optical spectra in the upper panels in Fig.~\ref{fig:stronglines}). Another strong line in the optical spectrum is a blend of \ion{He}{ii}~$\lambda6560\,\angstrom$ and hydrogen {\Halpha} (see zoom of the optical spectra in the lower panels in Fig.~\ref{fig:stronglines}). The strongest line in the UV spectra is \ion{He}{ii}~$\lambda$ $1640.5\,\angstrom$ while in the IR, we find the strongest line to be \ion{He}{ii}~$\lambda1.01\,\microm$ and \ion{He}{ii}~$\lambda1.86\,\microm$.
	
	At the CHeB stage, all models show almost only emission lines. These are much stronger than any emission line in the preceding CHB phases. 
	In addition to the \ion{He}{ii} lines, more metal lines of \ion{C}{\!} and \ion{O}{\!} begin to appear (see colored lines in upper panels in Fig.~\ref{fig:pMS-opt-D1-D10-r} and appendix Figs.~\ref{fig:pMS-uv-D1-D10-r} and \ref{fig:pMS-ir-D1-D10-r}). \ion{N}{\!} lines are again completely absent, except for \li{Ni}{V}{4606} but it is very weak. The strongest lines in the optical spectrum in this evolutionary phase are the oxygen doublet \ion{O}{vi}~$\lambda\lambda3811, 3834\,\angstrom$, carbon line \ion{C}{iv}~$\lambda4657\,\angstrom$ blended with \ion{He}{ii}~$\lambda4686\,\angstrom$, and 
\ion{C}{iv}~$\lambda7724\,\angstrom$. Additionally other lines are also strong, for instance \ion{O}{vi}~$\lambda4499\,\angstrom$, \ion{O}{vi}~$\lambda5288\,\angstrom$, and \ion{O}{vi}~$\lambda6191\,\angstrom$. In the UV region the strongest lines are \ion{O}{vi}~$\lambda1032\,\angstrom$ and doublet line \ion{C}{iv}~$\lambda\lambda1548, 1551\,\angstrom$, but also, \ion{O}{vi}~$\lambda1125\,\angstrom$, \ion{O}{vi}~$\lambda2070\,\angstrom$, and \ion{He}{ii}~$\lambda1641\,\angstrom$. In the IR region the strongest lines are \ion{O}{vi}~$\lambda1.08\,\microm$, \ion{O}{vi}~$\lambda1.46\,\microm$, and \ion{O}{vi}~$\lambda1.92\,\microm$.
	
	We infer that chemically-homogeneously evolving stars in early evolutionary phases show spectral features typical of weak and optically thin winds. Thus the term TWUIN star indeed applies for them. In the later evolutionary phases however, these stars become to exhibit spectral features which are common for stars with strong and optically thick winds. These are features typical of WR~stars. 
	
	Table \ref{tab:optdepth} lists the optical depths of the winds of our individual {\tt PoWR} models. We define these as the layers with a wind velocity of $v > 0.1\,$km s$^{-1}$. This is in line with the definition from Eq.\,(14) in \citet{Langer:1989a} applied in Paper I, but no longer explicitly relying on the $\beta$-law, although such a law is implicitly used in the atmosphere models. We list two different optical depth scales in Table \ref{tab:optdepth}: $\tau_{\mathrm{Thom}}$, which includes only the Thomson electron scattering, thereby allowing a direct comparison with the estimates made without a detailed atmosphere calculation in Paper I, while $\tau_{\mathrm{Ross}}$ is the Rosseland mean optical depth including all line and continuum opacity, which is an even more meaningful quantity for identifying optically thick regimes. We marked models having a wind optical depth of $\tau<1$ in both scales as TWUIN stars.
		
	We find that all models with reduced mass-loss rate (regardless of clumping) belong to TWUIN stars (i.e. they have a transparent wind) even in their CHeB stages. Models with nominal mass-loss rates and clumping develop an optically thick wind in their CHeB stages after having been experienced TWUIN phase during their CHB stages.
	%during the most of their lifetime have a transparent wind. The last one in their CHeB stages show optically thick winds while
	Additionally, the models with high $\mini$ (i.e. $\mini=59$ and $131$) have optically thick winds even in stages just before the CHeB phase (i.e. with $0.98\%$ of \ion{He}{\!}). {\it From this we can conclude that in the majority of their lifetimes, chemically-homogeneously evolving stars at low Z which emit most of their radiation in the UV (see Sec.~\ref{Ospectra})} have a transparent wind, i.e. they are TWUIN stars. However, the existence of several optically thick lines or continua in a part of the wind is not excluded.
	%Only in the CHeB stages this starts show optically thick winds, but the much longer CHB phase they have optically thin winds.
	All this is a consequence of the adopted {\mdot} prescriptions in the calculations, as we discuss in the following section. 

		\begin{table*}[!htbp]
        \centering
		\caption{Wind optical depths from our {\tt PoWR} models: $\tau_{\mathrm{Thom}}$ includes only electron scattering, while $\tau_{\mathrm{Ross}}$ includes all line and continuum opacity. Models which have $\tau<1$ in the wind in both of the $\tau$ scales, are marked as TWUIN star. Values of $\tau < 0.001$ are listed as zero.} 
			\begin{tabular}{lllllll}\hline\hline\small
		\rule[0mm]{0mm}{4.5mm}
			$D$ & $\mdot$ & M$_{\mathrm{ini}}$ & label & $\tau_{\mathrm{Thom}}$  & $\tau_{\mathrm{Ross}}$  &  TWUIN \\
			\hline
			\rule[0mm]{0mm}{4.5mm}
			%model      & $\log\mdot$  &  $D$   &  $\tau_{\mathrm{Thom}}$  & $\tau_{\mathrm{Ross}}$  &  TWUIN
& & 20 & T-1 (0.28) &  0.001  & 0.001 & YES \\
& & 20 & T-2 (0.50) &  0.002  & 0.002   & YES \\
& & 20 & T-3 (0.75) &  0.015  & 0.016   & YES \\
& & 20 & T-4 (0.98) &  0.222  & 0.235   & YES \\
& & 20 & T-5 (pMS)  &  0.914  & 1.112   & NO    \\ 
& & 59 &T-6 (0.28)  &  0.015  & 0.016   & YES \\
& & 59 &T-7 (0.50)  &  0.024  & 0.025   & YES \\
1 & higher& 59 &T-8 (0.75)&  0.149  & 0.157   & YES  \\
& & 59 &T-9 (0.98) &  3.740  & 4.613   & NO  \\
& & 59 &T-10 (pMS) &  4.456  & 5.748   & NO  \\ 
& & 131 &T-11 (0.28) &  0.059  & 0.062   & YES \\
& & 131 &T-12 (0.50) &  0.090  & 0.094   & YES \\
& & 131 &T-13 (0.75)&  0.589  & 0.622   & YES  \\
& & 131 &T-14 (0.98) &  3.460  & 3.848   & NO  \\
& & 131 &T-15 (pMS)  &  8.401  & 10.804  & NO  \\ \hline
& & 20 & T-1 (0.28) &  0.000  & 0.000  & YES \\  
& & 20 & T-2 (0.50) &  0.000  & 0.000  & YES \\ 
& & 20 & T-3 (0.75) & 0.000  & 0.000  & YES \\
& & 20 & T-4 (0.98) &  0.002  & 0.002   & YES \\ 
& & 20 & T-5 (pMS)  &  0.009  & 0.009   & YES \\ 
& & 59 &T-6 (0.28) &  0.037  & 0.041   & YES \\ 
& & 59 &T-7 (0.50) &  0.000  & 0.000  & YES \\ 
1 & reduced& 59 &T-8 (0.75) &  0.002  & 0.002   & YES  \\
& & 59 &T-9 (0.98) &  0.037  & 0.041   & YES \\ 
& & 59 &T-10 (pMS) &  0.048  & 0.054   & YES \\ 
& & 131 &T-11 (0.28) &  0.001  & 0.001 & YES \\ 
& & 131 &T-12 (0.50) &  0.001  & 0.001 & YES \\ 
& & 131 &T-13 (0.75) &  0.007  & 0.007   & YES  \\
& & 131 &T-14 (0.98) &  0.034  & 0.036   & YES \\ 
& & 131 &T-15 (pMS)  &  0.093  & 0.102   & YES \\ \hline
& & 20 & T-1 (0.28) &  0.001  & 0.001 & YES \\
& & 20 & T-2 (0.50) &  0.002  & 0.002   & YES \\
& & 20 & T-3 (0.75)&  0.015  & 0.016   & YES \\   
& & 20 & T-4 (0.98) &  0.225  & 0.256   & YES \\
& & 20 & T-5 (pMS)  &  0.926  & 1.054   & NO    \\ 
& & 59 &T-6 (0.28) &  0.014  & 0.015   & YES \\
& & 59 &T-7 (0.50) &  0.026  & 0.028   & YES \\
10& higher& 59 &T-8 (0.75)&  0.142  & 0.157   & YES  \\
& & 59 &T-9 (0.98) &  1.153  & 1.198   & NO  \\
& & 59 &T-10 (pMS) &  4.329  & 5.471   & NO  \\  
& & 131 &T-11 (0.28) &  0.057  & 0.061   & YES \\
& & 131 &T-12 (0.50) &  0.088  & 0.095   & YES \\
& & 131 &T-13 (0.75)&  0.649  & 0.673   & YES  \\
& & 131 &T-14 (0.98)  &  3.536  & 3.863   & NO  \\
& & 131 &T-15 (pMS)   &  8.447  & 10.776  & NO  \\ \hline
& & 20 & T-1 (0.28) &  0.000  & 0.000 & YES \\
& & 20 & T-2 (0.50) &  0.000  & 0.000 & YES \\
& & 20 & T-3 (0.75) &  0.000  & 0.000 & YES \\
& & 20 & T-4 (0.98) &  0.002  & 0.002   & YES \\
& & 20 & T-5 (pMS)  &  0.009  & 0.009   & YES \\ 
& & 59 &T-6 (0.28) &  0.000  & 0.000 & YES \\
& & 59 &T-7 (0.50) &  0.000  & 0.000 & YES \\
10& reduced& 59 &T-8 (0.75) &  0.002  & 0.002   & YES  \\
& & 59 &T-9 (0.98) &  0.015  & 0.015   & YES  \\
& & 59 &T-10 (pMS) &  0.042  & 0.047   & YES \\ 
& & 131 &T-11 (0.28) &  0.001  & 0.001 & YES \\
& & 131 &T-12 (0.50) &  0.001  & 0.001 & YES \\
& & 131 &T-13 (0.75)  &  0.007  & 0.008   & YES  \\
& & 131 &T-14 (0.98)  &  0.035  & 0.037   & YES \\ 
& & 131 &T-15 (pMS)   &  0.087  & 0.194   & YES \\\hline

		\end{tabular}
		\label{tab:optdepth}
	\end{table*}
%	\FloatBarrier
	
	%%%%%%%%%%%%%%%%%%%%%%%%%%%%%%%%%%%%%%%%%%%%%%%%%%%%%%%%
	\subsection{Effect of mass loss}\label{sec:masslossspectra}

	To study the effect of mass-loss rates on the synthetic spectra, we calculated a set of models with mass-loss rates that are hundred times lower than used in the stellar evolution calculations. Other parameters remained unchanged. These models are plotted as black lines in Figs.~\ref{fig:MS-opt-D1}, \ref{fig:stronglines}, and in the upper most panels of Figs.~\ref{fig:pMS-opt-D1-D10-r}, but also in Appendix Figs.~\ref{fig:MS-uv-D1}, \ref{fig:MS-ir-D1}, and in the upper most panels of Figs.~\ref{fig:pMS-uv-D1-D10-r} and \ref{fig:pMS-ir-D1-D10-r}.
	
	Models with lower mass-loss rate yield mostly absorption-line spectra during the CHB evolutionary phases. They show only negligible emission features (see the black lines in Fig.~\ref{fig:MS-opt-D1} and Appendix Figs.~\ref{fig:MS-uv-D1} and \ref{fig:MS-ir-D1}). Lowering the mass-loss rate affects the strength of the lines. While those few lines that are in emission become less intense, for most of those lines which are already in absorption using the lower mass-loss rates the absorption becomes even deeper. This illustrates that even pure absorption lines can be filled up by wind emission when applying higher $\dot{M}$. A more prominent effect of the same origin is the change of some lines from emission to absorption (see differences between colored and black lines in Fig.~\ref{fig:MS-opt-D1} and Appendix Figs.~\ref{fig:MS-uv-D1} and \ref{fig:MS-ir-D1}). 
   
    However, some absorption lines calculated with lower mass-loss rate become less pronounced in contrary what is expected as a general influence of lowering mass loss. A more prominent effect of the same origin (i.e absorption lines are switched to the emission) can be seen, for instance, in \ion{He}{ii}~$\lambda6560\,\angstrom$ blended with {\Halpha} (see the first and the second lower panels from the left in Fig.\ref{fig:stronglines}) and \ion{He}{ii}~$\lambda1.09\,\microm$, \ion{He}{ii}~$\lambda1.28\,\microm$, \ion{He}{ii}~$\lambda1.88\,\microm$, and \ion{He}{ii}~$\lambda1.88\,\microm$  blended with  \ion{H}{i} lines (see the upper two panels with orange and green boxes in Appendix Fig.~\ref{fig:MS-ir-D1}). These are due to a fact that for the models with higher percentage of hydrogen (more than $50\%$) \ion{He}{ii} lines which are in absorption are blended with hydrogen emission lines which are stronger. Combination of the \ion{He}{ii} absorption line and \ion{H}{i} emission lines results in the effect we described above. For more evolved models (i.e. which have much less or no helium at the surface) this effect is not visible. 
  
%	However, due to a higher abundance of hydrogen (more than $50\%$) most of the absorption lines are hydrogen lines a little bit influenced by helium. As a consequence, lines calculated with lower mass-loss rate are less intense (i.e. absorption is reduced) in contrary what is expected as general influence of lowering mass loss. A more prominent effect of the same origin (but absorption lines are switched to the emission) can be seen in {\Halpha} and IR lines \ion{He}{ii}~$\lambda1.28\,\microm$, \ion{He}{ii}~$\lambda1.86\,\microm$, and \ion{He}{i}~$\lambda1.88\,\microm$ (see for instance the first and the second lower panel in Fig.\ref{fig:stronglines} and the panels with orange and green boxes in appendix Fig.~\ref{fig:MS-ir-D1}). 
	
%	This effect may be seen in other observed stars as well. %\textbf{Other observed stars? Or stellar models? D.} 
%	If this is the case than it does not mean that lower mass-loss rates will always give less intense lines. Since the {\Halpha} line is the frequently used mass-loss diagnostics tool, the derivation of mass-loss based on this line can be inaccurate for chemically-homogeneously evolving stars, and should be double checked.
    
%    	The calculated synthetic spectra in earlier CHB phases (labeled with  $\YS=0.28$ and $\YS=0.5$) almost do not show any significant differences when reducing the mass-loss rate. 
	
	The low mass-loss spectra of less evolved TWUIN~stars (with $\YS=0.28$ and $\YS=0.5$) independent of their mass do not show any significant differences from their high mass loss counterparts except the effect we described above. Thus we can conclude that in early evolutionary stages, the assumptions about mass-loss in stellar evolutionary computations has a negligible effect, and would not lead to predicting different observable spectra. 
	
	At the evolutionary stage $\YS=0.75$, the spectra with $\mini=59$ and $131\,\msun$ show changes in some optical lines (e.g. \ion{He}{ii} at $\lambda4686\,\angstrom$ and $\lambda6560\,\angstrom$) from emission to absorption with lowering the mass-loss rate, while those with $\mini=20\,\msun$ still do not show any significant difference in their spectra (see the panel with the red box in Fig.~\ref{fig:MS-opt-D1}). Similar effect is seen in the UV and IR regions (see the panels with the red boxes in Appendix Fig.~\ref{fig:MS-uv-D1} and \ref{fig:MS-ir-D1}).
	
	The fact that chemically-homogeneously evolving stars in early evolutionary stages have weak and transparent winds, is in accordance with previous studies such as \citetalias{Szecsi:2015},
where authors where motivated to introduce the class of TWUIN stars.
	\FloatBarrier 
	For stars in the evolutionary stage $\YS=0.98$, the effect of the mass loss on spectra is more pronounced. These spectra show a few very weak emission lines, such as \ion{He}{ii} at $\lambda4200\,\angstrom$ and  \ion{He}{ii} $\lambda5412\,\angstrom$ (see the panel with the blue box in Fig.~\ref{fig:MS-opt-D1}). In the UV and IR spectral regions, the effect of lowering mass-loss rate is also visible (see the panels with the blue boxes in Appendix Fig.~\ref{fig:MS-uv-D1} and \ref{fig:MS-ir-D1}).
	The spectra with $20\,\msun$ show the same spectral features as more massive stars in previous evolutionary stages. 
	
	The most pronounced differences appear for the latest evolutionary stage. Indeed, models with CHeB show a strong dependence on the mass-loss rate applied, particularly for the stars with $\mini=59$ and $131\,\msun$ (see the uppermost panel in Fig.~\ref{fig:pMS-opt-D1-D10-r} and Appendix Figs.~\ref{fig:pMS-uv-D1-D10-r} and \ref{fig:pMS-ir-D1-D10-r}). These more evolved stars have a strong and thick wind with our default prescription, and thus lowering the mass-loss rates has an enormous influence on the resulting spectra. While the nominal mass loss produces very strong and broad emission features, the reduced one produces much less pronounced emission lines, if any. 
	
	We conclude therefore that, while varying mass-loss rates in the early evolutionary phases has no significant effect on the spectral appearance of these TWUIN~stars, having proper mass-loss rates for the more evolved stages where the models start to show WR-features is of uttermost importance. %For these stars even clumping factor may play an important role, as discussed below. %For that reason, we calculate third set of models assuming clumped wind.
	
	%{\bf Here we can write something about blending of \ion{H}{\!} and \ion{He}{\!} lines in IR part of the spectra. In this evolutionary stages \ion{H}{\!} lines are maybe in strong emission in IR part of the spectra ... In stages $\YS=0.25$ and $\YS=0.5$, \ion{H}{\!} is dominant ...}.
	%The strength of these lines increase from the less massive TWUIN~stars with $\mini=20~\msun$ to the more massive one with $\mini=131~\msun$. 
	%This suggests that in these evolutionary phases TWUIN~stars probably have no wind (or if they have a wind which has to be extremely weak). All spectral lines in these stages most probably have a photospheric origin. 
	
	\begin{figure*}[htp]\centering
		\includegraphics[width=1.28\textwidth,angle=90]{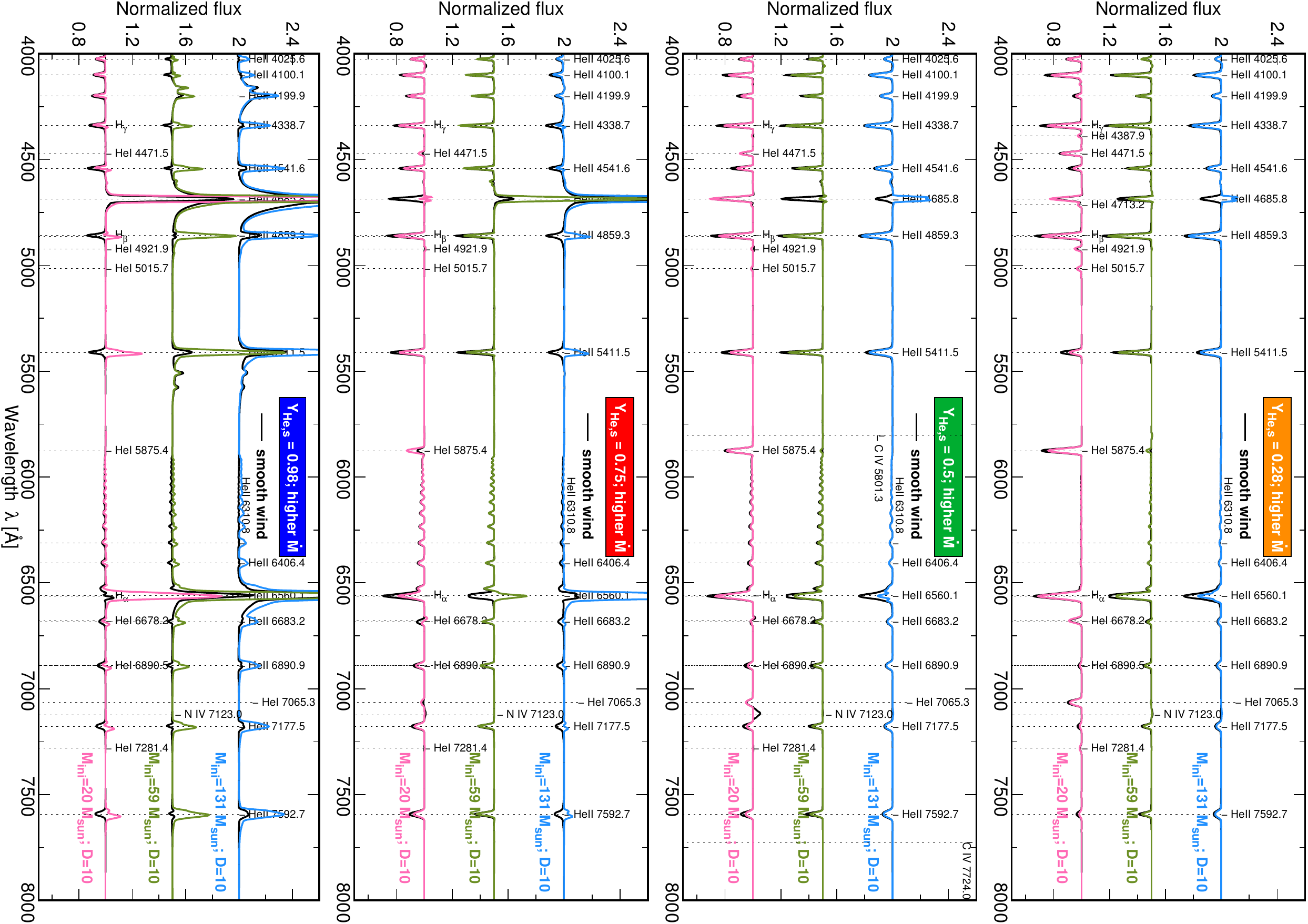}
		\caption{{\PoWR} spectra in the optical region of TWUIN~stars with different $\mini$ (see labels on the right side of the panels) and in different CHB evolutionary phases marked by the value $\YS$ in the colored boxes. The mass-loss rates are taken from the stellar evolutionary calculations (i.e. higher {\mdot}). Colored lines correspond to the clumped wind models (i.e. $D=10$), while the black lines correspond to the smooth wind.}
		\label{fig:MS-opt-D1-D10-hmd}
	\end{figure*}

	%\begin{figure*}[htp]\centering
	%\includegraphics[width=0.95\textwidth,angle=90,origin=c]{MS-opt.pdf}
	%\includegraphics[width=1.\textwidth]{MS-pMS-ir-D1-18-19.pdf}
	%	\caption{Zoom of the IR lines.}
	%	\label{fig:MS-pMS-ir-D1-18-19}
	%\end{figure*}

	\subsection{Effect of clumping}\label{sec:clumping}
	
	From observations and theoretical considerations we know that winds of almost all massive stars are inhomogeneous \citep[e.g.][]{Hamann:2008,Puls:2008}. The absence of direct observations of chemically-homogeneously evolving (TWUIN) stars means also that we do not have any observational constraint on clumping. However, we can check how wind inhomogeneities may influence the spectral appearance, from a purely theoretical point of view. Using a different clumping factor $D$, here we check how much the spectral appearance changes while keeping all other parameters the same.
	
	For our two sets of models, the one with the mass-loss rates as used in the stellar evolutionary models (higher {\mdot}) and the other with mass-loss rates hundred times lower (reduced {\mdot}), we calculated spectra with clumping factors $D=1$ (corresponding to a smooth wind) and $D=10$ assuming a clumping onset in the wind. 
	
	For the models with higher mass-loss rates the general influence of clumping on the spectral appearance is a reduction of absorption. As for the lines which are in emission in the smooth wind models, clumping makes them much more stronger. 
	Some lines even switch from absorption to emission, for instance the $\ion{He}{ii}~\lambda1641\,\angstrom$, $\lambda4686\,\angstrom$, $\lambda5412\,\angstrom$ lines and $\ion{He}{ii}~\lambda6560\,\angstrom$ line blended with the hydrogen $\Halpha~\lambda6563\,\angstrom$ line (see Fig.~\ref{fig:MS-opt-D1-D10-hmd} and also in Appendix Figs.~\ref{fig:MS-uv-D1-D10-hmd} and \ref{fig:MS-ir-D1-D10-hmd}).
	
	%Clumping has no significant influence on the spectral appearance of the models with $\mini=20$ and $59$ in earlier evolutionary phases (i.e. $\YS=0.28$ and $\YS=0.5$). The higher mass models in later evolutionary phases are more sensitive to clumping. 
	
	%In several models with $\YS=0.75$, some lines (e.g. \ion{He}{ii} $\lambda4685.78\,\angstrom$) turn from emission back to absorption. 
	%Synthetic spectra of models with higher mass-loss rates (colored lines in Figs.) are more sensitive to clumping than the models with lower mass-loss rates (black lines in Figs.).
	
	As for the models with reduced mass-loss rates, spectra during the CHB phases stay almost unchanged when clumping is taken into account (see comparison between colored and black lines in Appendix Figs.~\ref{fig:MS-opt-D1-D10-rmd}, ~\ref{fig:MS-uv-D1-D10-rmd}, and ~\ref{fig:MS-ir-D1-D10-rmd}). This effect is reasonable because with reduced mass-loss rates the wind becomes weaker, less dense, and more transparent. Hence, introducing clumping contributes very little to the changes of the optical properties of the wind. 
    
    The influence of clumping on spectral appearance is as expected. Models with the same $\mdot\sqrt{D}$ give similar spectra (at least the same equivalent with of the recombination lines). Therefore, if we increase clumping with the same {\mdot}, the spectra react as if we would increase {\mdot}. If {\mdot} is low enough not to affect the recombination lines very much we don't see much difference, which is why the low {\mdot} models don't show much difference.
    
%    The general influence of the clumping is a reduction of the absorption lines. However, the few lines which remain in emission in the smooth wind models become less intense (see $\ion{He}{ii}~\lambda6561.73\,\angstrom$ blended with the $\Halpha~\lambda6562.85\,\angstrom$ in the second panel from the top in Fig.~\ref{fig:MS-opt-D1-D10-rmd} and \ion{He}{ii}~$\lambda1.86\,\microm$ blended with \ion{He}{i}~$\lambda1.88\,\microm$ in appendix Fig.~\ref{fig:MS-ir-D1-D10-rmd}). 
	
\begin{figure*}\centering
		\includegraphics[page=1,angle=270,width=0.4999\columnwidth]{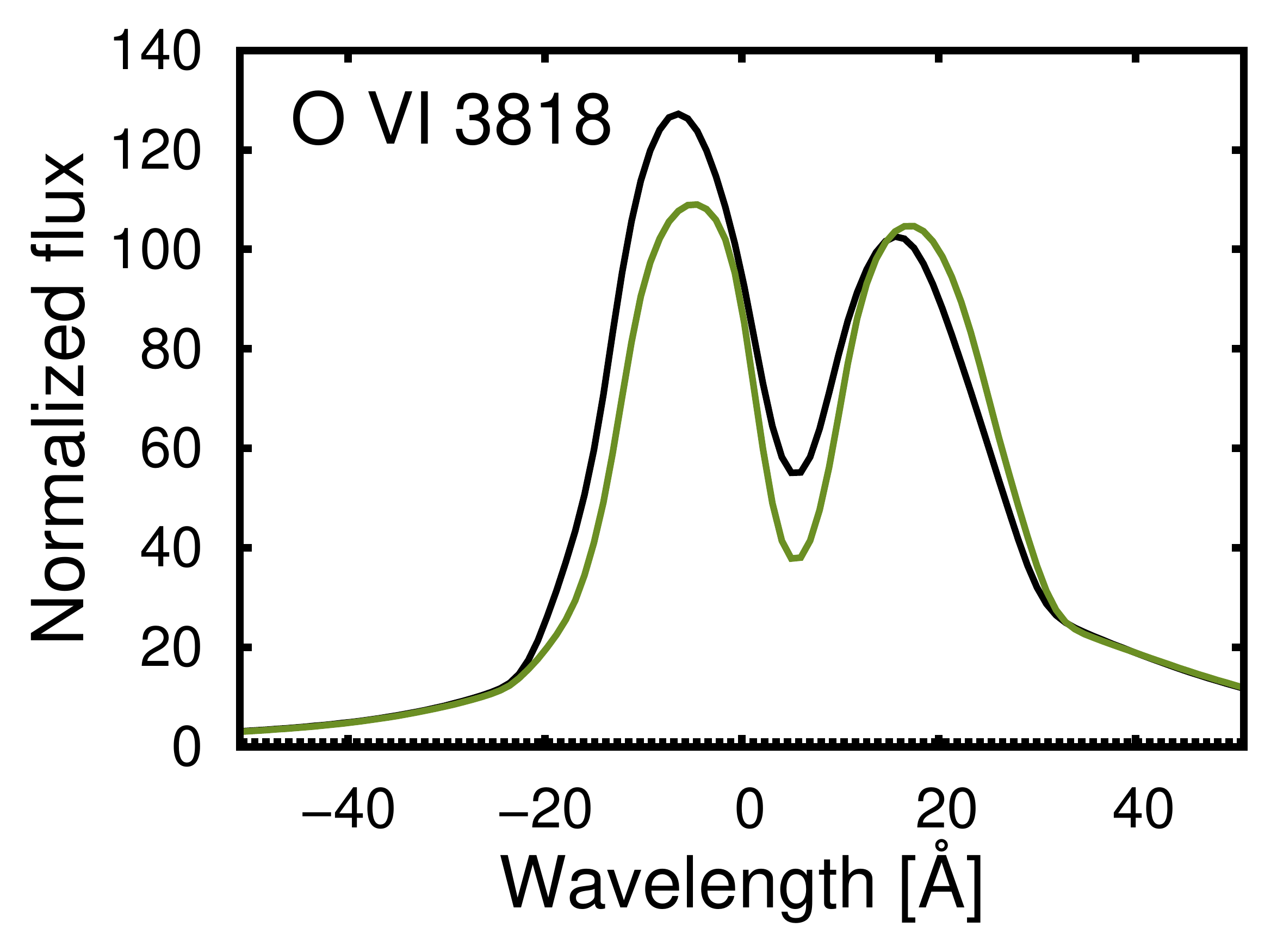}
		\includegraphics[page=2,angle=270,width=0.4999\columnwidth]{clumping-131-pMS.pdf}
		\includegraphics[page=3,angle=270,width=0.4999\columnwidth]{clumping-131-pMS.pdf}
		\includegraphics[page=4,angle=270,width=0.4999\columnwidth]{clumping-131-pMS.pdf}\\
		\includegraphics[page=5,angle=270,width=0.4999\columnwidth]{clumping-131-pMS.pdf}
		\includegraphics[page=6,angle=270,width=0.4999\columnwidth]{clumping-131-pMS.pdf}
		\includegraphics[page=7,angle=270,width=0.4999\columnwidth]{clumping-131-pMS.pdf}
		\includegraphics[page=8,angle=270,width=0.4999\columnwidth]{clumping-131-pMS.pdf}\\
		
\includegraphics[page=9,angle=270,width=0.4999\columnwidth]{clumping-131-pMS.pdf}
		\includegraphics[page=10,angle=270,width=0.4999\columnwidth]{clumping-131-pMS.pdf}
		\includegraphics[page=11,angle=270,width=0.4999\columnwidth]{clumping-131-pMS.pdf}
		\includegraphics[page=12,angle=270,width=0.4999\columnwidth]{clumping-131-pMS.pdf}        
        \caption{Influence of clumping on the line strength. Emission lines in the optical (upper panels), the UV (middle panels), and the IR (lower panels) regions of the model with $\mini=131~$M$_{\odot}$ in the CHeB evolutionary phase (mass-loss rate of $\log\zav{\mdot/\msunyr}=-4.23$). X-axis is centered around the wavelength indicated by the key legend (e.g. He~II~4686 means \ion{He}{ii}~$\lambda4686~\angstrom$). Green lines corresponds to clumped wind models with D$~$=$~10$, black to smooth wind models with D$~$=$~1$. %EW stands for equivalent width. 
		}\label{fig:importantlines}
	\end{figure*}

	Importance of clumping is more pronounced in the CHeB phase. In this stage the winds become stronger and denser and the contribution of clumping to the line formation becomes important. The models for CHeB stars with higher mass-loss rates show very pronounced emission lines, which become even stronger when clumping is taken into account, as seen in Fig.~\ref{fig:importantlines}. With clumping, the dense wind becomes more transparent, and thus more radiation can escape and contribute to the line strength (see the middle panels in Fig.~\ref{fig:pMS-opt-D1-D10-r}). However, the models with reduced mass-loss rates remain, even in this evolutionary stage, almost unchanged when clumping is taken into account (see the lowermost panel in Fig.~\ref{fig:pMS-opt-D1-D10-r}). A similar effect is seen in the UV and IR regions (see the middle and the lowermost panels in Appendix Figs.~\ref{fig:pMS-uv-D1-D10-r} and \ref{fig:pMS-ir-D1-D10-r}).

	%______________________________________________________________
	%                            ****
	%______________________________________________________________
	\section{Spectral classification}\label{sec:spectralclasification}
	
	We classify our model spectra according to the commonly used Morgan--Keenan spectroscopic classification scheme. We give a detailed description of this classification scheme in the context of hot massive stars in Appendix~\ref{sec:MK}.
	We report our findings summarized in Table~\ref{tab:classes}, and discuss some details below.
	
	%\newpage
	%\begin{landscape}
	\begin{table*}[t!]
    \centering
\caption{Spectral classification of our stellar models. TWUIN~stars (i.e. CHB objects) are typically assigned to some O-type subclass; a ``<''~sign marks if a model is consistent with earlier classes as well. For WR~stars, the spectrum may be consistent with more than one subclass; we give the secondary classification \citep[as defined in Table~3 of][]{Crowther:2006} in square brackets. See also the text and Appendix~\ref{sec:MK}.}
		%\vspace{20pt}
		\begin{tabular}{llllll}\hline\hline\small
		\rule[0mm]{0mm}{4.5mm}
			M$_{\mathrm{ini}}$ & label & $D=1$ & & $D=10$ & \\
			& & reduced $\mdot$ & nominal/higher $\mdot$ & reduced $\mdot$ & nominal/higher $\mdot$\\
			\hline
			 20 & T-1 (0.28) & O~8.5~V     & O~8.5~V    & O~9.5~V     & O~9~V      \\
 20 & T-2 (0.5)  & O~5.5~III   & O~6~III    & O~7~III     & O~7~III      \\
 20 & T-3 (0.75) & <\,O~4~III  & <\,O~4~III & <\,O~4~III  & O~5~I     \\
 20 & T-4 (0.98) & O~4~III     & <\,O~4~I   & <\,O~4~III  & O~4~I      \\
 20 & T-5 (pMS) & WO~2 [–]    & WO~1     & WO~2 [–]    & WO~1 [WO~3]     \\\hline
 59 & T-6 (0.28) & <\,O~4~III     & <\,O~4~III     & <\,O~4~III     & <\,O~4~III      \\
 59 & T-7 (0.5) & <\,O~4~III     & <\,O~4~III     & <\,O~4~III     & <\,O~4~I      \\
 59 & T-8 (0.75) & O~4~III     & <\,O~4~I     & <\,O~4~III     & <\,O~4~I      \\
 59 & T-9 (0.98) & <\,O~4~III    & <\,O~4~I     & <\,O~4~III     & [WO~2~or~WO~1]      \\
 59 & T-10 (pMS) & WO~1 [WO~3]    & WO~1     & WO~1 [WO~3]    & WO~1      \\\hline
 131 & T-11 (0.28) & O~4~III     & <\,O~4~III     & <\,O~4~III     & <\,O~4~I      \\
 131 & T-12 (0.5) & O~4~III     & <\,O~4~III     & <\,O~4~III     & <\,O~4~I      \\
 131 & T-13 (0.75) & <\,O~4~III     & <\,O~4~I     & O~4~III     & <\,O~4~I      \\
 131 & T-14 (0.98) & <\,O~4~III     & O~4~I     & <\,O~4~III     & WO~4 [WO~2~or~WO~1]     \\
 131 & T-15 (pMS) & WO~1 [WO~3]    & WO~1     & WO~1 [WO~3]    & WO~1      \\ \hline

		\end{tabular}
		%\\ $\star$ means Fe2-10
		\label{tab:classes}
	\end{table*}
	%\end{landscape}
	
	\subsection{TWUIN stars are very hot O\,stars}
%	\subsection{\bf CHB TWUIN stars appear as early-type O stars}
	Most of our spectra that show almost no emission lines, that is, the stellar models that have been designated as TWUIN~stars in \citetalias{Szecsi:2015}, are assigned to the class O\,4 or earlier. It means they are very early O-type giants or supergiants. This is because the logarithm of the ratio of \li{He}{I}{4473} to \li{He}{II}{4543} being smaller than $-$0.6 makes them belong at least to type~O\,4 \citep{Mathys:1988}, and in the absence of nitrogen lines we cannot distinguish between earlier classes \citep[as done e.g. in][]{Walborn:2002}. In fact, the ratio of said helium lines is usually around -1.5 or less. Thus, all we can safely say for these stars is that they are of class O~4 or earlier.
	
	Luminosity classes for the spectra which are consistent with classes earlier than O~4 type (marked as <\,O~4 in Table~\ref{tab:classes}) are defined based on the nature of the line \li{He}{II}{4686}. If it is found in emission, the spectrum is classified as a supergiant (i.e. luminosity class~I). If it is found in weak absorption \citep[i.e. logarithm of the absolute value of the equivalent width being lower than 2.7, cf.][]{Mathys:1988}, the spectrum is classified as a giant (i.e. luminosity class~III) and if strongly in absorption, a dwarf (i.e. luminosity class~V).
	
	We find late type O~stars, that is, O\,5 to O\,9.5, only among the lowest mass models (with M$_{ini}$~$=$~20~M$_{\odot}$). As for their luminosity classes, we apply two criteria: one for those earlier than O\,8 as explained above, and another for those between O\,8.5$-$O\,9.5 (cf. Appendix~\ref{sec:MK}). This other criterion is provided by \citet[][]{Conti:1971} and based on the equivalent width ratio of the lines \li{Si}{IV}{4090} and \li{He}{I}{4143} \citep[but see also][]{Martins:2018}.
	With this, our spectra of a 20~M$_{\odot}$ star are assigned to dwarf (V) at the ZAMS and to giant (III) in the middle of the MS~phase. However, this distinction seems to be an artifact of using two different criteria for those earlier and later than O\,8. As Fig~\ref{fig:HRD} and Table~\ref{tab:list} attest, the radius of the 20~M$_{\odot}$ model does not change significantly between the phases {\YS}~$=$~0.28 and {\YS}~$=$~0.5. The luminosity does change however, showing that the conventional nomenclature associated with luminosity classes (giant, dwarf etc.) may not always be very meaningful in accounting for the radial size of a star. 
	
	We did not find any of our spectra to be consistent with the O\,f subclass \citep{Crowther:1995,Crowther:2011}. This is because the defining feature of this subclass, the line \li{N}{III}{4640}, is completely absent in all our spectra. The O~f subclass practically means that the star has a fairly strong wind; this is why galactic early-type stars tend to have it. But it is not surprising that our low-metallicity stars with weak winds do not show this feature.
	
	Some of our <\,O\,4 stars are really hot. \citet{Tramper:2014} investigated ten low-metallicity (down to 0.1~\Zsun) O-type stars and found the hottest to be T$_{\rm eff}$~$=$45~kK, while our hottest O-type object has T$_{\rm eff}$~$=$85~kK. 
	\textit{Thus the detection of a very hot, early O-type star at low metallicity without an IR-excess would be a strong candidate for a star resulting from chemically homogeneous evolution.} We refer to our Sect.~\ref{sec:PoWR-TWUIN-comp} where we compare one of our <\,O~4 type spectra to a regular O~type stellar spectra from the literature.

	\subsection{TWUIN stars turn into Wolf--Rayet stars in the CHeB phase}
	
The term Wolf--Rayet~stars refers to a spectral class, based on broad and bright emission lines observed in the optical region. As mentioned shortly in Sect.~\ref{sec:Intro}, authors working on stellar evolution sometimes refer to objects that are hot and (more or less) hydrogen-deficient as WR~stars, too. From an evolutionary point of view, a massive star's surface can become hydrogen-poor because of the (partial) loss of the hydrogen-rich envelope either by Roche-lobe overflow \citep[a scenario originally suggested by][]{Paczynski:1967}, or by stellar winds \citep{Conti:1975}. A third option that can lead to a hydrogen-deficient surface composition is internal mixing (e.g. due to rotation, as in the present work). Nonetheless, the fact that a stellar model's surface is hydrogen-poor, does not necessarily mean that its wind is optically thick \citepalias[as shown by Sect.~6 of][]{Szecsi:2015}. It also does not mean that broad emission lines develop (as shown by our CHB spectra), although they may (as shown by our CHeB spectra). Below we discuss the spectral classes of the latter case.
	
	All our spectra of the CHeB phase show features typical for WR~stars of the WO~type: strong \li{C}{IV}{5808}, \li{O}{V}{5590} and \li{O}{VI}{3818} in emission. We classify these objects according to criteria in Table~3 of \citet{Crowther:1998}. There are two main criteria, a primary and a secondary. We find that sometimes these two do not provide the same class. % or the secondary is not applicable. 
    In this case we mention the secondary classification in square brackets in our~Table~\ref{tab:classes}. 
    
    %We find no oxygen emission lines in the CHeB models of 20~M$_{\odot}$ 
    %with reduced mass-loss rate; but the presence of the strong \li{C}{IV}{5808} emission line makes them consistent with the class WC~4.
    %{\color{violet}
    %Curiously, we find our most massive terminal-age main-sequence models with nominal mass-loss rate to be neither consistent with the O, nor with the WC/WO sequences. This is because they do not show \li{C}{IV}{5808}, nor oxygen in emission, which would make them an early WC or a WO~type star; but also they show \li{He}{II}{4543} in emission which excludes them from the O~type sequence too. \li{He}{II}{4686} is also strongly in emission in them. However, such a star may not be expected to be actually observed, since this phase is already close to the end of CHE stage and thus possibly a short-lived phase. 
    %}
	
	We find that nitrogen lines are almost completely absent. The line \li{N}{V}{4606} is sometimes present, most of the time in absorption. When it is in emission, its equivalent width never increases above 0.3~$\angstrom$ which means it is very weak. Other lines typical for WN~stars \citep{Smith:1996} such as \li{N}{III}{4640} and \li{N}{IV}{4057} are not found in any of our spectra. 
	The almost complete absence of N-lines may make a future observer consider such a star to be some other type, certainly not WN. 
	
	Thus we conclude that after first producing very hot early O~type stars during the CHB phase, chemically homogeneous evolution leads to WO~type stars during the CHeB phase. We remind that the CHeB lifetime is about 10\% as long as the CHB lifetime. Therefore, in a population of chemically-homogeneouly evolving stars, we expect to find 10~times more hot early O~stars than WO~stars.

    % I put the table to the beginning of this section (D.)
%______________________________________________________________
	\section{Discussion}\label{sec:discussion}
	
	\subsection{Comparison to synthetic spectra from the literature}\label{sec:PoWR-TWUIN-comp}
	
	\subsubsection{O~type spectra}\label{Ospectra}
\begin{figure*}\centering
			\includegraphics[width=1.3\textwidth,angle=90]{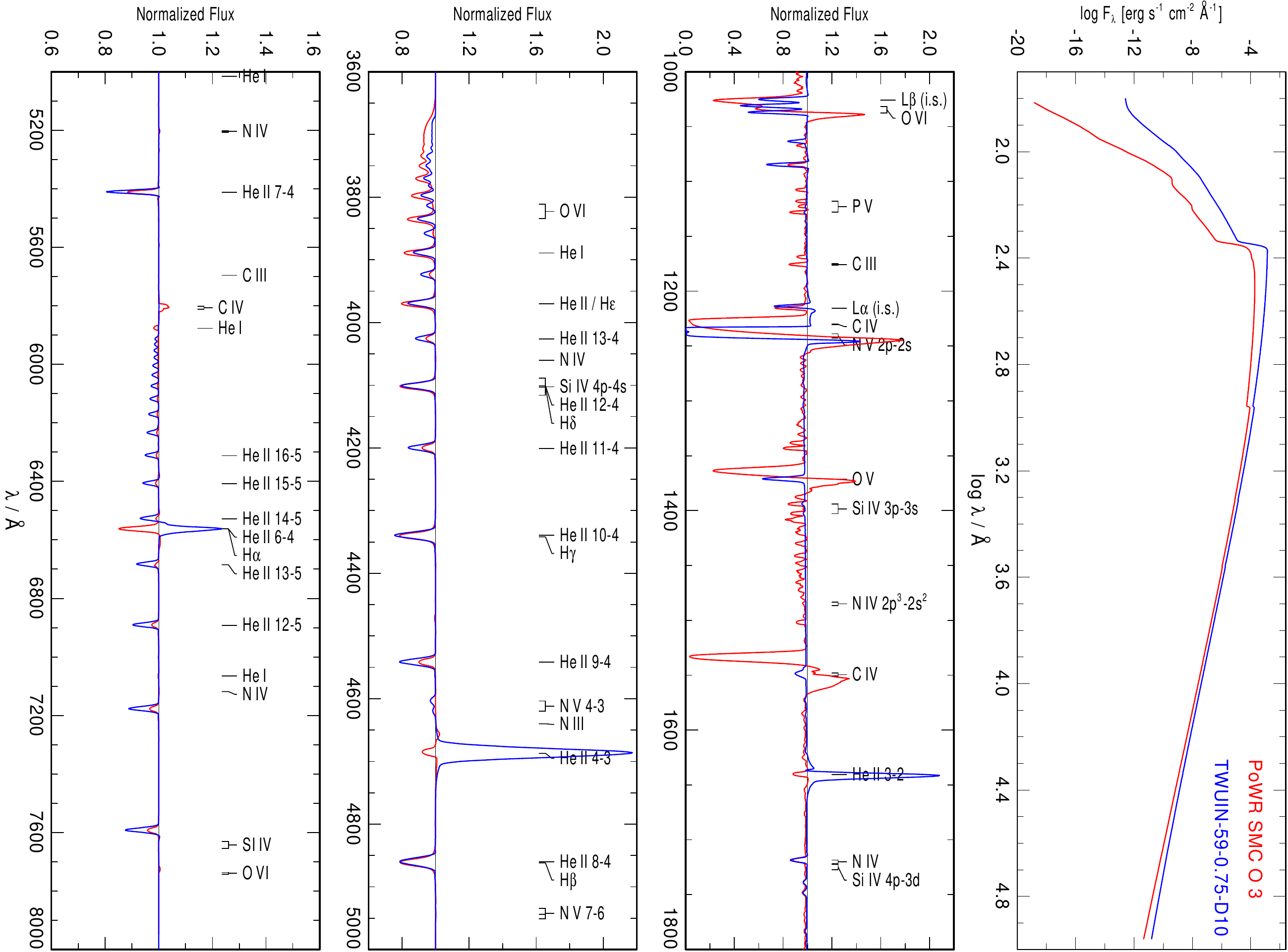}
			\caption{
				Comparison of one of our representative O\,3~III spectra (i.e. TWUIN T-8 model, see Table~\ref{tab:list}) to an O\,3 synthetic star from the literature. Both stars are placed at the same distance of 10 pc. 
				An effect of line broadening due to fast rotation is taken into account as mentioned in Sec.~\ref{sec:spectralmodels} assuming the same rotational velocity which corresponds to the T-8 model. See Sect.~\ref{sec:PoWR-TWUIN-comp} for more details. 
			}\label{fig:PoWR-TWUIN-comp}
		\end{figure*}

	We compare one of our absorption line spectra to a typical O\,type spectra in the literature. In particular, we do the comparison between our model labeled T-8 in Table~\ref{tab:list} (with clumping and nominal mass-loss rate), and a model spectrum of an O\,3 star taken from the \PoWR ~SMC OB model grid \citep[see][]{Hainich:2018arXiv} database\footnote{http://www.astro.physik.uni-potsdam.de/$\sim$wrh/PoWR/SMC-OB-II/} (see Fig.~\ref{fig:PoWR-TWUIN-comp}) corresponding to the composition of the Small Magellanic Cloud (Z$_{\mathrm{SMC}}$~$\sim$~0.2~Z$_{\odot}$). 
	The parameters of the SMC~O\,3 model are: $\Teff=50~\kKelvin$, $\log\zav{\lstar/\lsun}=5.52$, $\log\zav{\mdot/\msunyr}=-5.9$, $\log\zav{g/\cmss}=4.4$, and $D=10$. Note that the main differences between the two models are (i) the metallicity (ours is about ten times lower while \ion{He}{} abundance is almost the same) and (ii) the surface temperature (ours is 69~\kKelvin). An effect of smoothing and broadening the lines due to fast rotation are taken into account (see Sec.~\ref{sec:spectralmodels}) for both models assuming the same rotational velocity which corresponds to the T-8 model (see Table~\ref{tab:list}).
	%, and (iii) stellar rotation (ours is a fast rotating star). This last fact has an effect of smoothing and broadening the lines compared to the non-rotating star's spectra.
	
	From comparing the SEDs of both stars placed at the same distance of 10~pc (see the uppermost panel in Fig.~\ref{fig:PoWR-TWUIN-comp}) we can infer that the amount of emitted far and extreme UV ionizing radiation increases  particularly at shorter wavelengths (around the \ion{H}{i} ionization edge). This is consistent with the fact that our model has a higher surface temperature. Another effect which may lead to higher UV flux is that there is less line blanketing at low-metallicity, therefore less flux is being redistributed to longer wavelengths.
	
	As for the spectral features, we can infer the following. In the optical region, the SMC~O\,3 spectrum shows the \ion{C}{iv}~$\lambda\lambda5801, 5812\,\angstrom$ lines while in the TWUIN T-8 model we do not find any metal lines. %In the optical region, the SMC~O\,3 spectrum shows some metal lines (for instance \ion{N}{iv}~$\lambda4060\,\angstrom$, \ion{C}{iii}~$\lambda5696\,\angstrom$, and \ion{C}{iv}~$\lambda\lambda5801, 5812\,\angstrom$) while in the {\bf TWUIN T-8} model spectra we do not find any of these metal lines. 
	In the UV region, the SMC~O\,3 spectra also shows very strong metal lines (for instance doublet \ion{O}{vi}~$\lambda\lambda1032, 1038\,\angstrom$, doublet \ion{N}{v}~$\lambda\lambda1239, 1243\,\angstrom$, \ion{O}{v}~$\lambda1371.3\,\angstrom$, and doublet \ion{N}{iv}~$\lambda\lambda1548, 1551\,\angstrom$) which are not present in the TWUIN T-8 model spectra. This is expected because the TWUIN~star models have very low metallicity. 
	
	\ion{He}{ii} lines are in very strong emission in the TWUIN T-8 model spectra (for instance \ion{He}{ii}~$\lambda1641\,\angstrom$, \ion{He}{ii}~$\lambda4687\,\angstrom$, and \ion{He}{ii}~$\lambda6562\,\angstrom$) while in the SMC~O\,3 spectra those lines are in absorption. This is consistent with the fact that the TWUIN model has a high surface helium abundance ({\YS}$\sim$0.5). On the other hand,
	\ion{He}{i} lines are not present in the TWUIN T-7 model spectra while in SMC~O\,3 they are visible (see \ion{He}{i}~$\lambda5877\,\angstrom$, \ion{He}{i}~$\lambda7065\,\angstrom$, and \ion{He}{i}~$\lambda3888\,\angstrom$
lines in Fig.~\ref{fig:PoWR-TWUIN-comp}). 

\subsubsection{WO~type spectra}

We compare our emission line spectra to a typical WO~type spectra from the literature.
Comparing our models to a WO~star model was not an easy task, as
there exist no analyses of observations of WO~stars with the metallicity we study here, and, consequently, no models as well. As for somewhat higher metallicities such as Z$_{\rm SMC}$, very few models have ever been calculated. Here we use a model from \cite{Shenar:2016}, which was applied for the analysis of the SMC binary star AB\,8 with following model parameters: $\Teff=141~\kKelvin$, $\log\zav{\lstar/\lsun}=6.15$, $\log\zav{\mdot/\msunyr}=-4.8$, $\vinf=3700~\kms$, $\log g \zav{\cmss}=5.4$, and $D=40$. These parameters resemble our T-10 model (see Table~\ref{tab:list}) with clumping and nominal mass-loss rate.

\begin{figure*}\centering
			\includegraphics[width=1.3\textwidth,angle=90]{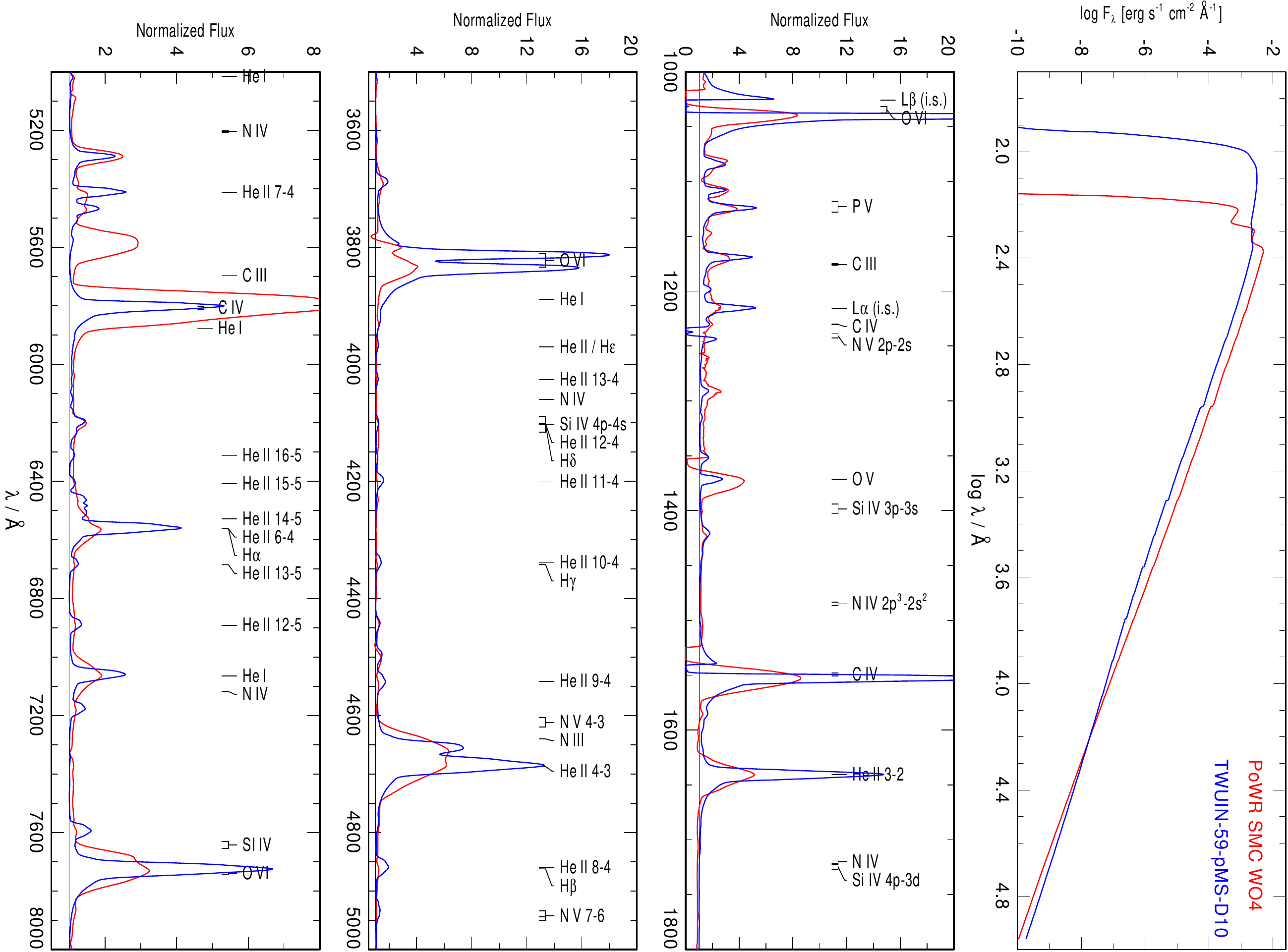}
			\caption{Same as Fig.~\ref{fig:PoWR-TWUIN-comp} but comparing our WO~1 spectra (i.e. TWUIN T-10 model, see Table~\ref{tab:list}) to an SMC WO\,4 synthetic spectrum from \cite{Shenar:2016}.}\label{fig:TWUIN-WR-comp}
		\end{figure*}

The comparison is shown in Fig.~\ref{fig:TWUIN-WR-comp}. For both models rotation is taken into account assuming the same rotational velocity which corresponds to the T-10 model (see Table~\ref{tab:list}). The \mbox{T-10} model is somewhat more luminous. Both models have the same beta (i.e. $\beta=1$), similar mass-loss rates and $\log g$. The mass-loss rates of both models are relatively high, but we do expect the Z-dependency to drop for WO stars, because their atmospheres are enriched with fusion products that are not related to the initial metallicity, and these products contribute to the driving of their winds. Also, the WO star in the SMC seems to be much more luminous than the Galactic ones, which increases its mass-loss rate compared to other WO stars.

The main differences between the two models are (i) the surface temperature (ours is 138~\kKelvin), (ii) the terminal velocity (ours is 1000~\kms), (iii) clumping factor (ours is 10), and (iv) the metallicity and element ratios (see mass fractions in Table~\ref{tab:list}). The mass-fraction of the WO~4 model are: \ion{He}{\!}=0.399, \ion{C}{\!}=3$\cdot10^{-1}$, \ion{O}{\!}=3$\cdot10^{-1}$, and \ion{Fe}{\!}=6$\cdot10^{-4}$. In the WO\,4 model \ion{N}{\!} was not included while in the T-10 model it is. For both models \ion{H}{\!} is not included. The T-10 model has about twice as much, somewhat less \ion{C}{\!}, by more than two orders of magnitude less \ion{O}{\!}, and by more than one order of magnitude less iron group elements. 

The difference in SED (see uppermost panel in Fig.\,9) can be attributed to the different \ion{Fe}{} abundances in the models. The \ion{Fe}{} abundance in the WO star model is more than an order of magnitude
higher than in our T-10 model, causing substantial absorption and reemission of UV photons in the visual part (line blanketing). The difference in the spectral line shapes can be mainly attributed to
differences in $v_\infty$, which is more than a factor three larger in the WO model. Finally, the large differences in the strength of some spectral lines is a result of differences in two things: the abundances, and the so-called "transformed radii" $R_{\rm t}$, which represent an integrated
emission measure in the wind (see equation\ 1 in \citealt{Hamann:2006}). The SMC WO model has a higher $R_{\rm t}$ value, and hence overall weaker spectral lines. Although the spectra of these models differ significantly, they do predict similar lines appearing in the spectrum.

To conclude, the different environment and formation history of chemically-homogeneous evolving stars could mean that they appear somewhat different than the SMC WO component at their evolved phases. Their exact appearance would depend on parameters such as the terminal velocity, which was fixed in our study. Regardless of this uncertainty, however, we find that so-called TWUIN stars appear as WO stars in their evolved phases.

	\subsection{On the validity of our model assumptions}\label{sec:validity}
	
	Our wind models and emergent spectra are theoretical predictions based on current knowledge of stellar evolution and stellar wind modelling. However, they are also subject to several assumptions.
	
	The radial wind velocities in our models were assumed to follow the $\beta$-law in Eq.~\eqref{betavel}. Although there exist several calculations of the wind velocity law which take into account acceleration of matter by scattered and absorbed radiation either in an approximate way using force multipliers \citep[e.g.][]{Castor:1975, Abbott:1980, Pauldrach:1986} or in a more exact way using
	detailed radiative transfer \citep[e.g.][]{Abbott:1982, Grafener:2005, Pauldrach:2012, Krticka:2010, Krticka:2017, Sander:2017}, the $\beta$-velocity law became a standard assumption in modelling stellar wind spectra.
	Using a free parameter $\beta$ allows to find a velocity law which fits the observations best, regardless the consistency of such result.
	As discussed by \citet{Krticka:2011}, the $\beta$-velocity law is not a bad approximation to consistent hydrodynamical calculations, however, there exists a better and more exact fit using Legendre polynomials.
	In any case, using the $\beta$-velocity law is a reasonable first approximation.
	
	We also had to assume a mass-loss rate. This was done by simply taking the mass-loss rates {\em assumed} in the stellar evolutionary models. As explained in Sect.~\ref{sec:massloss} the evolutionary models assumed a mass-loss rate following certain recipes. The recipe of
	\citet{Vink:2000, Vink:2001} in Eq.~\eqref{eq:vink} was derived from atmosphere simulations that used detailed treatment of a full line list for a fixed velocity law. The recipe of \citet{Hamann:1995} given in Eq.~\eqref{eq:hamann} was, on the other hand, based on observed spectra of WR~stars. 
	Nonetheless, these prescriptions may not be valid when it comes to TWUIN~stars. Thus, to make sure that we understand the consequences of using them anyway, we tested the effect of lowering the mass loss by one hundred, to be consistent with the findings of \citet[][cf. our Table~\ref{tab:massloss}]{Hainich:2015}. The results of this test in the context of line formation was reported in Sect.~\ref{sec:masslossspectra}. %Further studies are needed to establish the correct mass-loss rates of TWUIN stars both observationally and theoretically. 
	 
	 %\textcolor{purple}{To test how our lower mass-loss rates agree with the s. We also derived multidimensional fits to express the mass-loss rate as a function of the stellar parameters} 
	
	%From Sect.~\ref{sec:masslossspectra} we conclude that for the early stages of evolution, mass loss has very little influence on the spectra. Only in the later phases of their evolution, in particular during CHeB, we found strong effects. The same is true not only for the line spectrum but for the spectral energy distributions as well.
	
	As chemically-homogeneously evolving stars are generally fast rotators ($\sim$0.6~\vcrit) this may have some yet unexplored effect on their wind structures \citep{Owocki:1996}. However, this would require detailed hydrodynamical calculations in at least 2D, which are beyond the scope of current paper. We accounted for the spectral imprint of fast rotation by performing a flux-convolution of the emergent spectrum with a rotation profile. 
	
	Here we only studied the spectra of single stars, but a large fraction of massive stars are born in close binary systems and thus undergo interaction with a companion during the evolution at some point \citep[e.g.][]{Paczynski:1967,Sana:2012,Gotberg:2017}. However, the ratio of binary stars vs. single stars is unknown at the metallicity we study, and may be quite different from the Galactic case (e.g. because the stability of the collapsing star-forming cloud may be influenced by its metallicity).
    For example, TWUIN~stars in a close binary orbit have been suggested to be the stellar progenitors of compact object mergers explaining the origin of gravitational waves \citep{Marchant:2016,Marchant:2017}. 
    How such an interaction with a companion influences the spectral appearance remains to be studied.

	\subsection{Future research on TWUIN~stars -- theory}\label{sec:theory}
	
	Taking the same mass-loss rate that was assumed when computing the evolution, makes our spectral predictions consistent with evolutionary models.
	However, in the absence of actual observations of TWUIN~stars, it is a question whether such a star can have a wind at all. 
	Testing this can be done similarly to how it was done by
	\citet{Krticka:2014} for the case of winds with non-solar CNO~abundances. Although this test is computationally expensive and goes beyond the scope of current work, here we summarize the basic idea, as well as the results we may expect from such a test, as a motivation for future work.
	
	As mentioned, we assumed that the wind structure of all TWUIN~stars can be described by a $\beta$-law \citep[e.g.][]{Puls:2008}, motivated by e.g. hydrodynamical consistent calculations for Galactic WN~stars \citep{Graefener:2008}. Additionally, we assumed input parameters ($\beta$, \vinfty) that are typical for hot massive O-type stars and WR~stars at $0.2\dots1\,Z_\odot$. All this may not hold for extremely low-metallicity environments; and the issue is further complicated by the observed steep metallicity dependence of the mass loss found by \citet{Hainich:2015} as well as the so-called ``weak wind problem'' \citep[see e.g.][]{Martins:2005, Marcolino:2009, Huenemoerder:2012}.
	%described by \citet{Huenemoerder:2012}.
	
	One way to validate our assumptions used in this work would be hydrodynamical simulations of the wind and its structure. This has been done for Galactic massive O\,stars in \citet{Krticka:2017} and a few WR~stars in \citet{Graefener:2008}. Although expensive, such simulations for the models presented in this work could provide essential information on how valid our spectra's predictions are. 
	
	For example, if atmospheres models based on hydrodynamic simulations point to different values for $\beta$, $\dot{M}$, or {\vinfty}, this will influence the predicted line strengths in our spectra and thus lead to assigning different spectral classes for these stars. The models may even show that the $\beta$-law as such is not applicable at all in this regime or that these stars -- at least during some parts of their evolution -- might not have winds at all. Thus, future studies in this direction are highly needed. 
	
	As for metallicity, here we only applied one set of stellar evolutionary models, all computed with Z$_{ini}$~$=$~0.02~Z$_{\odot}$. However, chemically homogeneous evolution is predicted to happen at various sub-solar metallicities \citep[see e.g.][]{Brott:2011a}. Although its prevalence is expected to be larger when lowering the metallicity (see Sect.~10.4 of Paper\,1), it is nonetheless an important future research direction to study the spectra of chemically-homogeneously evolving stars up to at least Z$_{\mathrm{SMC}}$. 
	
	\subsection{Future research on TWUIN~stars -- observations}\label{sec:observations}
	
	It is essential to obtain observational samples of metal-poor massive stars to test our theories. Ideally, we would need an extensive spectral catalog of about 50-100 massive stars at metallicities lower than 0.1~\Zsun. This task seems challenging, but not at all impossible with the most modern observing facilities and the next generation telescopes coming up. For example, ESO's MUSE spectrograph can take optical spectra of several dozens of massive stars in local-group galaxies \citep{Castro:2018}, while systematic studies of these spectra (including spectral classification and determination of mass-loss rates) could be done with state-of-the art tools \citep[e.g.][]{Hillier:1998,Puls:2005,Gustafsson:2008,Kamann:2013,Tramper:2013,Ramachandran:2018}.
	
	Until we obtain a comprehensive census of \textit{individual} massive stars at low-metallicity, we may compare our predictions to observed \textit{populations} of massive stars. Such a comparison of our model predictions to unresolved observed features of massive star populations in the dwarf galaxy I~Zwicky~18 \citep{Kehrig:2015,Kehrig:2016} is planned in a subsequent work. 
	
	Another interesting application of our spectra could be done in the context of the reionization history of the Universe. It has been suggested that massive stars and, especially, chemically homogeneous evolution may be important for this process \citep[e.g.][]{Eldridge:2012}, as WR-like emission bumps are often \textit{not} found in the spectra of high-redshift galaxies. Note that our spectral models suggests that TWUIN~stars are indeed not expected to show prominent emission lines because their winds are rather weak. Thus, these stars' contribution to the reionization epoch should be also investigated in the future.

	%({\bf should we describe here results from J.Krti\v cka calculation for the star T61kK-M130-L6.41-Md-7.3-v966-H51-D1-Rot924 for which he obtained $\vinf=966\,\kms$ and $\beta=0.8$}). 
	
	%More research is needed to determine if this simple approach is valid; in particular as the fast rotation ($\sim$0.4$-$0.6~$v_{\mathrm{break-up}}$) may lead to an asymmetric wind. But this approach is the best what we can do using 1-D stellar atmosphere models.
	%______________________________________________________________
	\section{Summary and conclusions}\label{sec:conclusion}
	
	We studied the spectral appearance of chemically-homogeneously evolving stars, as predicted by evolutionary model sequences of fast rotating massive single stars with low metallicity.
	To compute spectra, we employed the NLTE model stellar atmosphere and stellar wind code \PoWR. We predicted detailed spectra for selected stars from three evolutionary models, namely those with initial mass $20$~$\msun$, $59$~$\msun$, and $131$~$\msun$. Various evolutionary stages were studied (comprising the CHB and CHeB phases).
	The stellar parameters, namely the effective temperature,
	luminosity, mass, and chemical composition, were taken from the evolutionary
	models. Wind models and their spectra were calculated for fixed values of the
	terminal velocity and velocity law. We tested the influence of two of the most uncertain assumptions in stellar wind modelling, mass-loss rate and clumping. The model spectra were classified according to the Morgan--Keenan spectroscopic classification scheme. Our main findings are summarized below:
	
	\begin{itemize}
		\item Our models in early evolutionary phases have weak and optically thin winds while in later phases these stars exhibit stronger and optically thick winds. This is consistent with earlier studies (see \citetalias{Szecsi:2015} who established that in the early phases these objects should be called TWUIN~stars), and is a consequence of the adopted {\mdot} prescription. When adopting a reduced mass-loss rate, we find only a few weak emission lines in the spectra even in the most evolved phases.
		
		\item The maximum of the emitted radiation is in the far and extreme UV region. The emitted radiation in the \ion{He}{ii} continuum increases both with {\mini} and the evolutionary status, later stages having higher emissions. The total emitted flux is not very sensitive to variations of either the mass-loss rate or clumping. 
		
		\item In earlier evolutionary phases with 50\% of hydrogen or more in the atmosphere, most of our spectra show, independently of their {\mini}, almost exclusively absorption lines. This is true for the whole spectral region. More emission lines start to appear in later evolutionary phases, shortly before the end of the CHB phase. In the CHeB phase almost all lines are found in emission. In particular, helium emission lines are strong and very characteristic for evolved stars. Their line strengths increase with higher helium abundance.
		
		\item Our models predict that 
		lower mass-loss rates than those
		adopted from the evolutionary calculations have negligible effect on the emergent spectra of the TWUIN~star models in early evolutionary phases. More pronounced influence on spectral appearance is seen in later evolutionary phases with more helium in the atmosphere; especially in the CHeB phase. 
		%In this evolutionary phase having exact value of the mass-loss rates in the stellar evolutionary calculation is much more important.
		
		\item Assuming clumped wind has no significant influence on the predicted TWUIN spectra in earlier evolutionary phases. The spectra of higher mass models in later evolutionary phases are, on the other hand, sensitive to clumping. Reducing the mass-loss rate however cancels out this sensitivity, that is, model spectra with reduced mass-loss rates remain almost unchanged when assuming clumped wind, even in late evolutionary phases.
		
		\item Our TWUIN model spectra are assigned to the spectral class O\,4 or earlier. Nitrogen lines are almost completely absent. TWUIN O-type stars are predicted to be much hotter than those O-type stars that have been so far observed spectroscopically (down to 0.1~Z$_{\odot}$). Thus, the detection of a very hot star without almost any metal lines but with strong \ion{He}{ii} emission lines that is consistent with some very early type~O giant or supergiant, would be a strong candidate for a star resulting from chemically homogeneous evolution. 
        
        \item In later evolutionary phases most of our model spectra are assigned to the WO-type spectral class. Nitrogen lines are almost completely absent in this late phase as well. Thus, chemically-homogeneous evolution first leads to very hot early type O~stars (TWUIN~stars) and then, for the last $\lesssim$~10\% of the evolution, to Wolf--Rayet stars of type~WO.
        
        \item The fact that chemically homogeneously evolving stars only develop emission lines during their CHeB phase, but have only absorption lines during their long lived CHB phase (when they are TWUIN~stars), suggests that these stars may have contributed to the reionization of the Universe. Observations of high-redshift galaxies typically show that an intensive ionizing source is present that produces almost no WR-like emission bumps in the galactic spectra. Some populations of TWUIN~stars may be this source. 
		
		%As our model estimates indicate even lower mass-loss rates than those adopted from the evolutionary calculations, we calculated another set of
		%spectra with a mass-loss rates two orders of magnitude lower.
		%Resulting spectra were almost identical, so the low mass-loss rate has a
		%negligible effect on the emergent spectra.
		
		%\item In later evolutionary stages with more helium in the atmosphere, wind emission lines with P-Cyg type profiles appear as a consequence of a denser wind.
		
		%Generally, the spectra of evolved TWUIN~stars are characterized by stronger helium absorption lines. Their strengths rises with higher helium abundance. For stars with higher mass some lines turn into an emission for higher helium abundance. The final model for CHeB phase (without hydrogen) shows emission lines for all three selected evolutionary sequences.
	\end{itemize}
	
	Single stars with cheically-homogeneous evolution may be the progenitors of long-duration gamma-ray bursts and type~Ic supernovae, as shown by e.g. \citet{Yoon:2006} and \citet[][in Chapter~4]{Szecsi:2016}. In a close binary system, they may lead to two compact objects that eventually merge, giving rise to detectable gravitational wave emission \citep{Marchant:2016,Marchant:2017}. Indeed, our choice of metallicity was motivated by the fact that at this metallicity, binary models predict a high rate of gravitational wave emitting mergers.
	
    As our test with the two mass-loss rate values indicate, even if the mass-loss rate turns out to be much lower than what is applied in the evolutionary models during the CHB phase -- and indeed even if some of these stars turn out not to have winds at all -- our conclusions about the absorption-like spectra will remain the same. As for the CHeB phase, there the mass-loss rate plays an important role; we suggest to carry out hydrodynamic simulations of the wind's structure for these stars, to be able to constrain their mass-loss rates and thereby to investigate their spectral appearance further.
	
    Due to the lack of spectroscopic observations of individual massive stars with metallicity below 0.1~\Zsun, we could not compare our spectra with observations of any stars that may be of similar nature. Indeed, the main purpose of our work is to motivate future observing campaigns aiming at low-metallicity starforming galaxies e.g. \textit{Sextant~A} or \textit{I~Zwicky~18}.
    
    %Another indirect evidence for the existence of TWUIN~stars is their ability to account for the high ionizing emission in starforming dwarf galaxies. Yet, we do not have any direct observational clue about their lives. We hope that our theoretical models can be tested against observations of individual massive stars in the near future, for example in dwarf galaxy \textit{I~Zwicky~18} or other star-forming galaxies of low metallicity.
	
	\begin{acknowledgements}
		This research was supported by the grant 16-01116S (GA \v{C}R).
		The Astronomical Institute Ond\v{r}ejov is supported by the
		project RVO:67985815. 
		%{\bf B.~K. would like to thank R.~Hainich for  valuable discussion.}
        D.~Sz. is thankful for the relevant discussion to G.~Gr\"afener and N.~Langer, as well as to \'A.~Szab\'o as usual. 
		A.A.C.S. is supported by the Deutsche Forschungsgemeinschaft (DFG) under grant HA 1455/26 and would like to thank STFC for funding under grant number ST/R000565/1. TS acknowledges funding from the European Research Council (ERC) under the European Union’s DLV-772225-MULTIPLES Horizon 2020 research and innovation program.
	\end{acknowledgements}

	%-------------------------------------------------------------------
	%%%%%%%%%%%%%%%%%%%% REFERENCES %%%%%%%%%%%%%%%%%%
	
	\bibliographystyle{aa} % style aa.bst
	\bibliography{References} % your references Yourfile.bib
	%%%%%%%%%%%%%%%%%%%%%%%%%%%%%%%%%%%%%%%%%%%%%%%%%%
	
	%%%%%%%%%%%%%%%%% APPENDICES %%%%%%%%%%%%%%%%%%%%%
	% * <brankica.kubatova@asu.cas.cz> 2017-10-27T20:14:22.796Z:
	%
	\begin{appendix}
		
		\onecolumn
		
		% The chapter on the details of how we classified is moved from the text to the appendix. 
		\section{Spectral classification}\label{sec:MK}
		
		The Morgan--Keenan spectroscopic classification scheme is based on comparing the strengths of certain lines. That is, if the ratio of two given lines falls into an (observationally pre-defined) regime, the star is assigned to a certain class. For example, if the ratio of the lines \li{He}{I}{4473} to \li{He}{II}{4543} falls between, say, 0.2 and 0.1, the spectra is classified as type O\,8. 
		
		The line strength is usually measured by the equivalent width of the line. Typically in the literature, the ratio of two lines is expressed as the logarithm\footnote{When we talk about logarithm, we always mean log$_{\rm 10}$ unless specified otherwise.} of the ratio of their equivalent widths\footnote{The equivalent width \textsl{ratio} is sometimes denoted as log$_{\rm 10}$~$W_{\lambda}$ in the literature. We warn about this notation being contradictory, as also the equivalent width \textsl{itself} is commonly denoted by log$_{\rm 10}$~$W_{\lambda}$.}, that is, log$_{\rm 10}$~$(EW_{\rm line1} / EW_{\rm line2})$. %In the present work, when the line is in absorption, we define the EW as negative; when in emission, positive. Since we can only make sense of the ratio of two lines if both are in absorption or both are in emission (since for taking the logarithm the ratios should always be positive), applying this convention has no effect on our classification. 
		
		For O-type stars, the work of \citet{Mathys:1988}, who in turn relied on the works done by \citet{Conti:1971,Conti:1974,Conti:1977}, defines subclasses comprehensively. They take into account the ratio of \li{He}{I}{4473} to \li{He}{II}{4543} when defining the spectral subclasses between type~O~3 (early) to O~9.7 (late); the classification scheme we base our present work is given in Table~III of \citet{Mathys:1988}. \citet{Walborn:2002} updated this scheme for the earliest types, introducing type~O~2; however, they use the ratio of certain nitrogen lines, which are absent from our spectra. 
		Additionally, in e.g. paragraph~6 of Sect.~4.2 of \citet{Mathys:1988}, O\,f subclasses are defined on the basis of the \li{N}{III}{4640} line; this line is also absent from our spectra. 
		
		As for luminosity classes of O\,type stars, we classify everything with \li{He}{II}{4686} in emission as a supergiant, (i.e. luminosity class~I). For dwarfs (class~V) and giants (class~III) on the other hand, \citet{Mathys:1988} suggests the following approach. For spectral types earlier than O\,8.5 (that is, types between O\,3--O\,8), he uses the line strength of \li{He}{II}{4686} to distinguish between luminosity classes. His criterion is given in paragraph~4 of his Sect.~4.2: if strongly in absorption, meaning log~$|\mathrm{EW}|$~$>$~2.7, it is of class~V (note the absolute values). If only weakly in absorption, it is of class~III. 
		For spectral types O\,8.5 and later, he uses the \textsl{sum} of the logarithm of two lines, \li{He}{I}{4388} and \li{He}{II}{4686}.  %this criterion is given in his Table~IV. 
		However, we found that in our spectra both of these lines are too weak, so even their sum is not an applicable criterion. Instead, we rely on \citet{Conti:1971} for these late spectral types, who use the equivalent width ratios of \li{Si}{IV}{4090} to \li{He}{I}{4143} with a criterion given in their Table~5. 
		
		For WR~stars, we have to distinguish between so-called nitrogen-sequence WR~stars (type~WN) on the one hand, and carbon- and oxygen-sequence WR~stars (WC and WO) on the other. 
		
		WN stars are typical in that they have strong nitrogen emission lines---in particular, \li{N}{III}{4640} and \li{Ni}{IV}{4059} \citep[][]{Crowther:1995,Smith:1996,Crowther:2011}. Also, \li{He}{II}{4686} is in emission in their spectra. There is a comprehensive set of criteria for WN~classification in Table~4a of \citet{Smith:1996}. According to this table, we find no WN stars amongst our spectra. 
		%Smith:1996 define spectral classes based on \textit{peak line to continuum ratios} which is practically the maximum of the line in a normalized flux vs. wavelength plot. I guess?
		
		Quantitative classification of WC and WO\,stars has been done by \citet{Crowther:1998}. In their Table~3, equivalent width ratios of certain carbon- and oxygen-lines are used to distinguish between classes from WC~11 to WC~4, as well as from WO\,4 to WO\,1. We rely on this system for classifying those spectra that have strong emission features in carbon and oxygen. Note however that the line \li{C}{III}{5696} which is used to distinguish between WC~type subclasses is completely absent from our spectra, leading us to classify all our emission line spectra into type~WO.

		%%%%%%%%%%%%%%%%%%%%%%%%%%%%%%%%%%%%%%%%%%%%%%%%
		
		% HERE COMES THE SPECTRA IN APPENDIX
		
		%%%%%%%%%%%%%%%%%%%%%%%%%%%%%%%%%%%%%%%%%%%%%%%%		
		
		\section{Spectral models of TWUIN~stars}\label{sec:appendix}
		
		\renewcommand\thefigure{B\arabic{figure}}    
		%\setcounter{figure}{0}

		%  OPTICAL REGION
		
\begin{figure*}
\centering
	\includegraphics[width=0.99\textwidth]{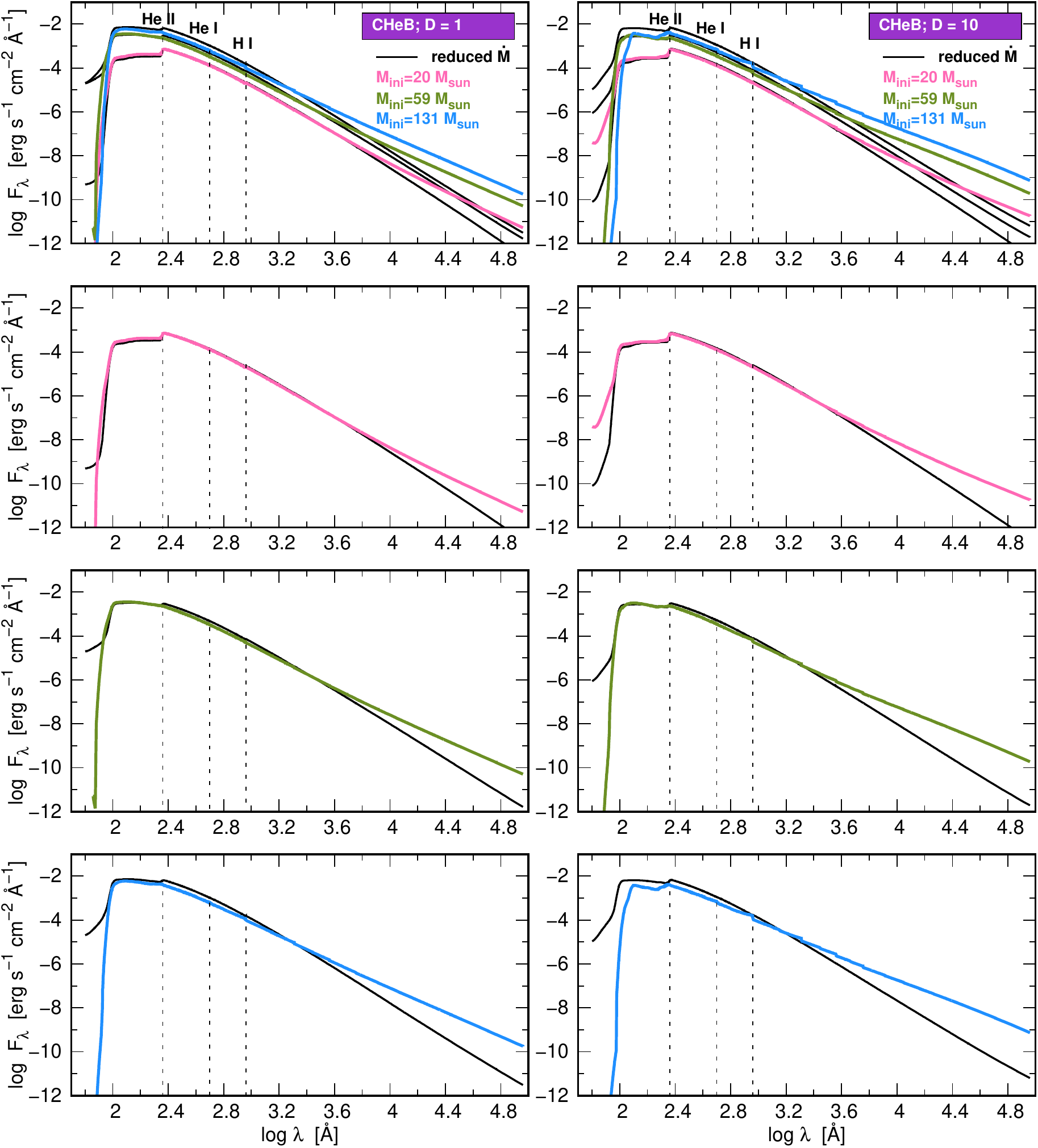}
			\caption{Same as Fig.~\ref{fig:sed-evol} but only for CHeB evolutionary phase.}\label{fig:SED-D1-D10-evol-cheb}
		\end{figure*}	
        
\begin{figure*}
\centering        
  \includegraphics[width=0.96\textwidth]{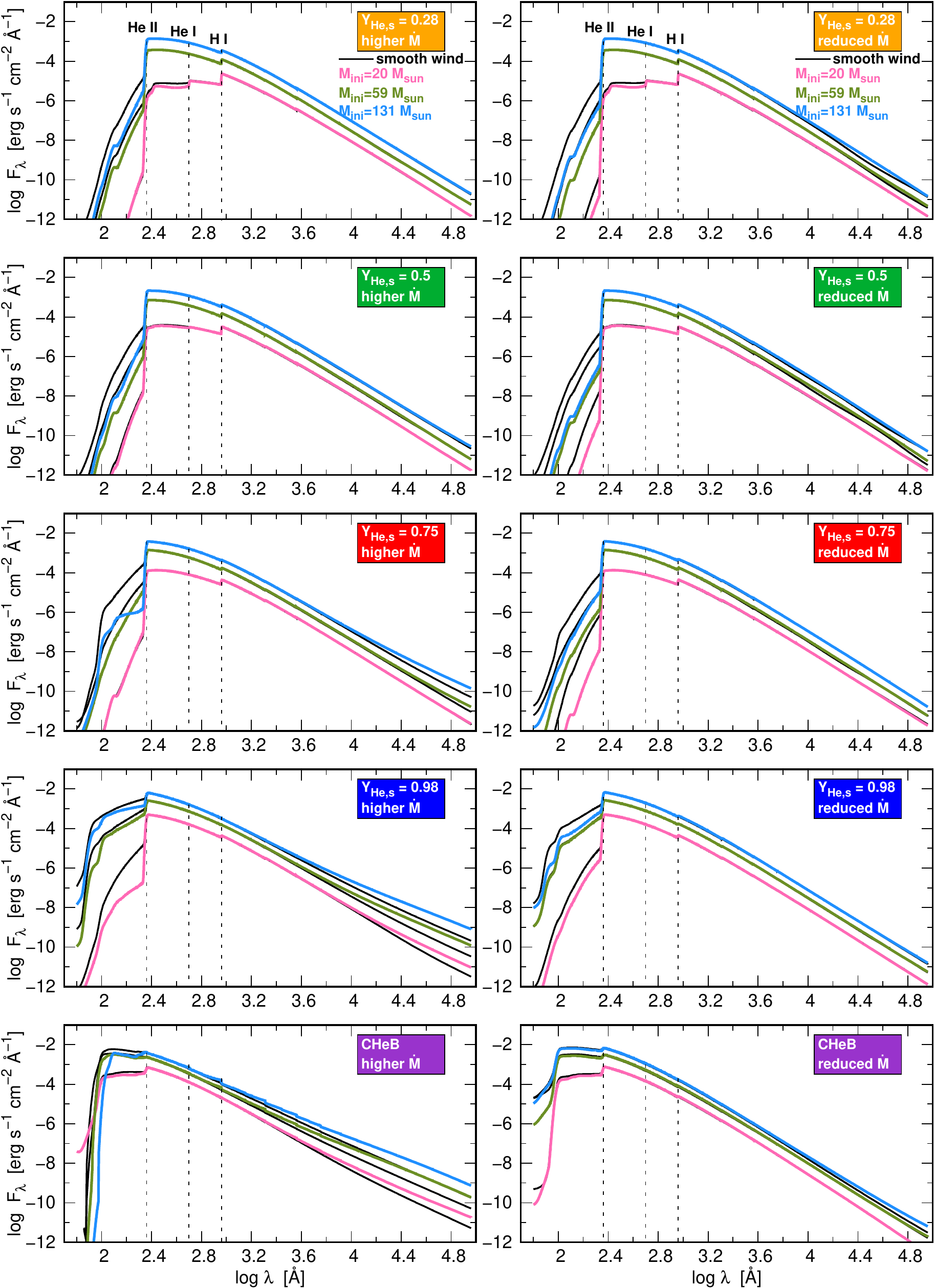}
			\caption{Same as Fig.~\ref{fig:sed-evol} but differences between SEDs are shown for smooth (black lines) and clumped (colored lines) wind models for higher (left panels) and reduced (right panels) mass-loss rate {\mdot}.}\label{fig:SED-D1-D10-evol-clump}
		\end{figure*}      
		
		%%%%%%%  UV REGION  MS D = 1 %%%%%%%%%%%%%%%%%%%%%%%%%%%%%%%%%%%%%%%%%%%%%%%%%%%%%%
		
		\begin{figure*}\centering
			\includegraphics[width=1.28\textwidth, angle =90]{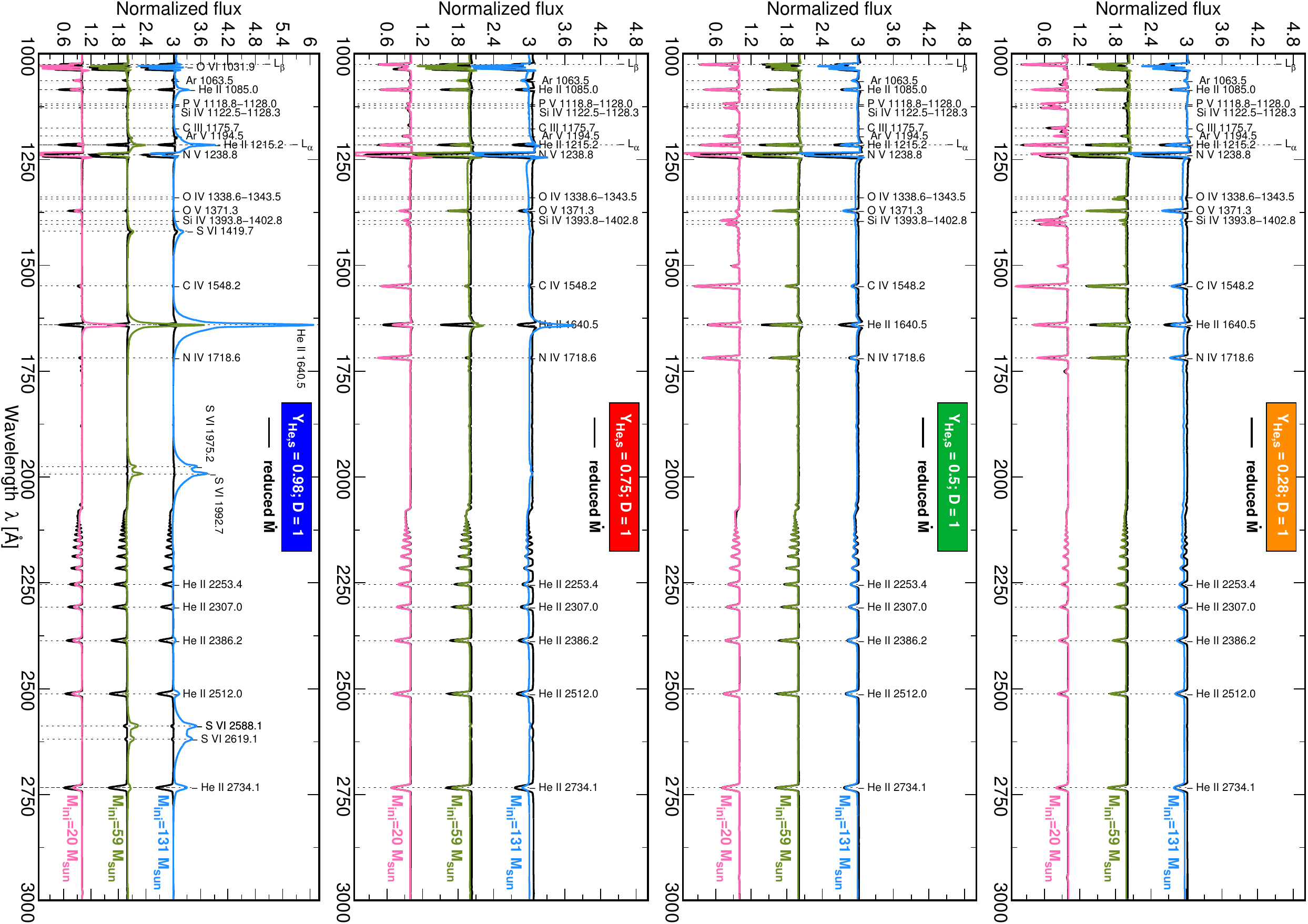}
			\caption{Same as Fig.~\ref{fig:MS-opt-D1} but in the UV region.}\label{fig:MS-uv-D1}
		\end{figure*}
		
		%!!!!!! pMS D = 1 and D = 10. !!!!!!!
		\begin{figure*}\centering
			\includegraphics[width=0.9\textwidth, angle =90]{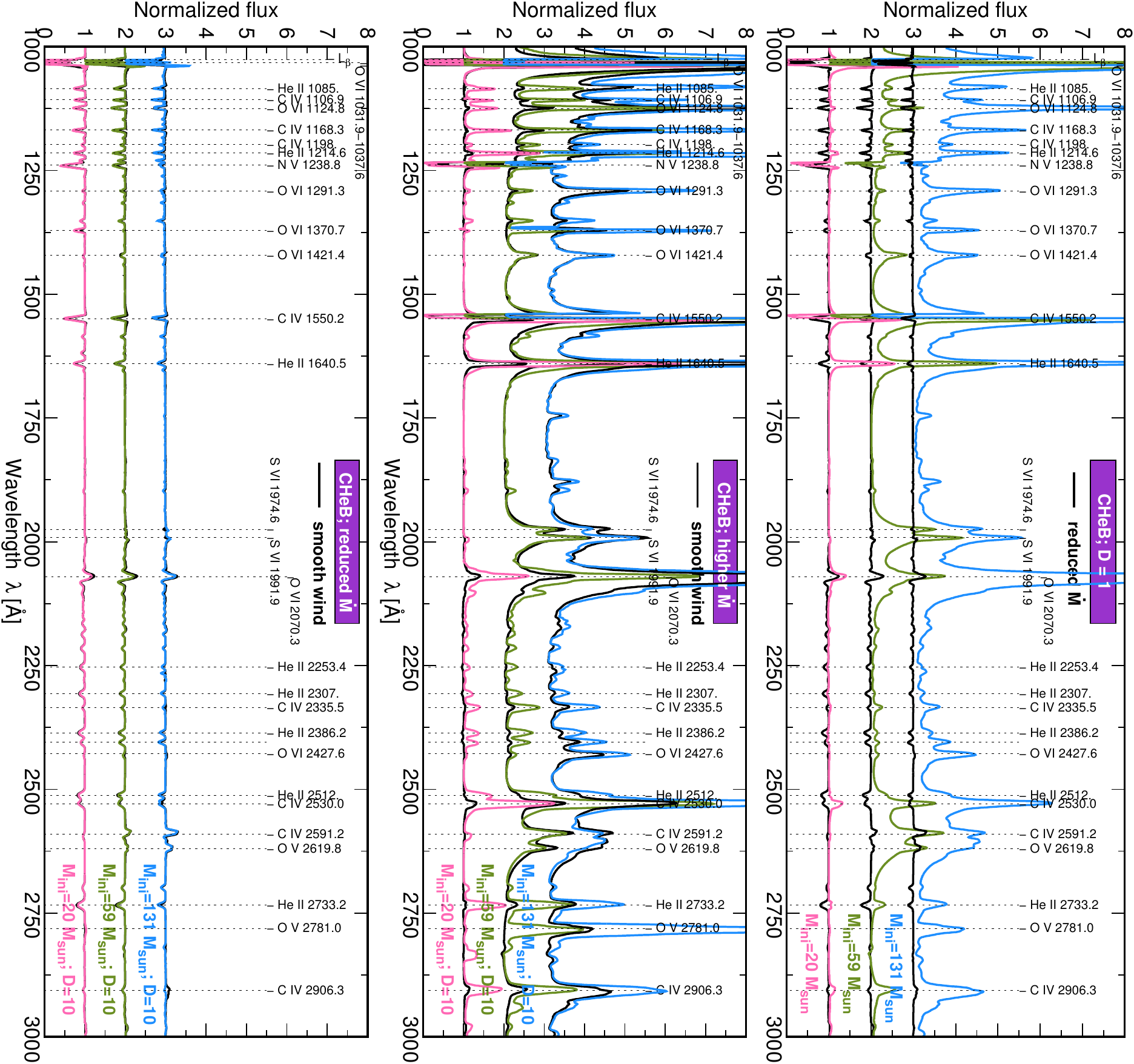}
			\caption{Same as Fig.~\ref{fig:pMS-opt-D1-D10-r} but in the UV region.}\label{fig:pMS-uv-D1-D10-r}
		\end{figure*}
		
		%%%%%%%  IR REGION  MS D = 1  %%%%%%%%%%%%%%%%%%%%%%%%%%%%%%%%%%%%%%%%%%%%%%%%%%
		
		\begin{figure*}\centering
			\includegraphics[width=1.3\textwidth, angle =90]{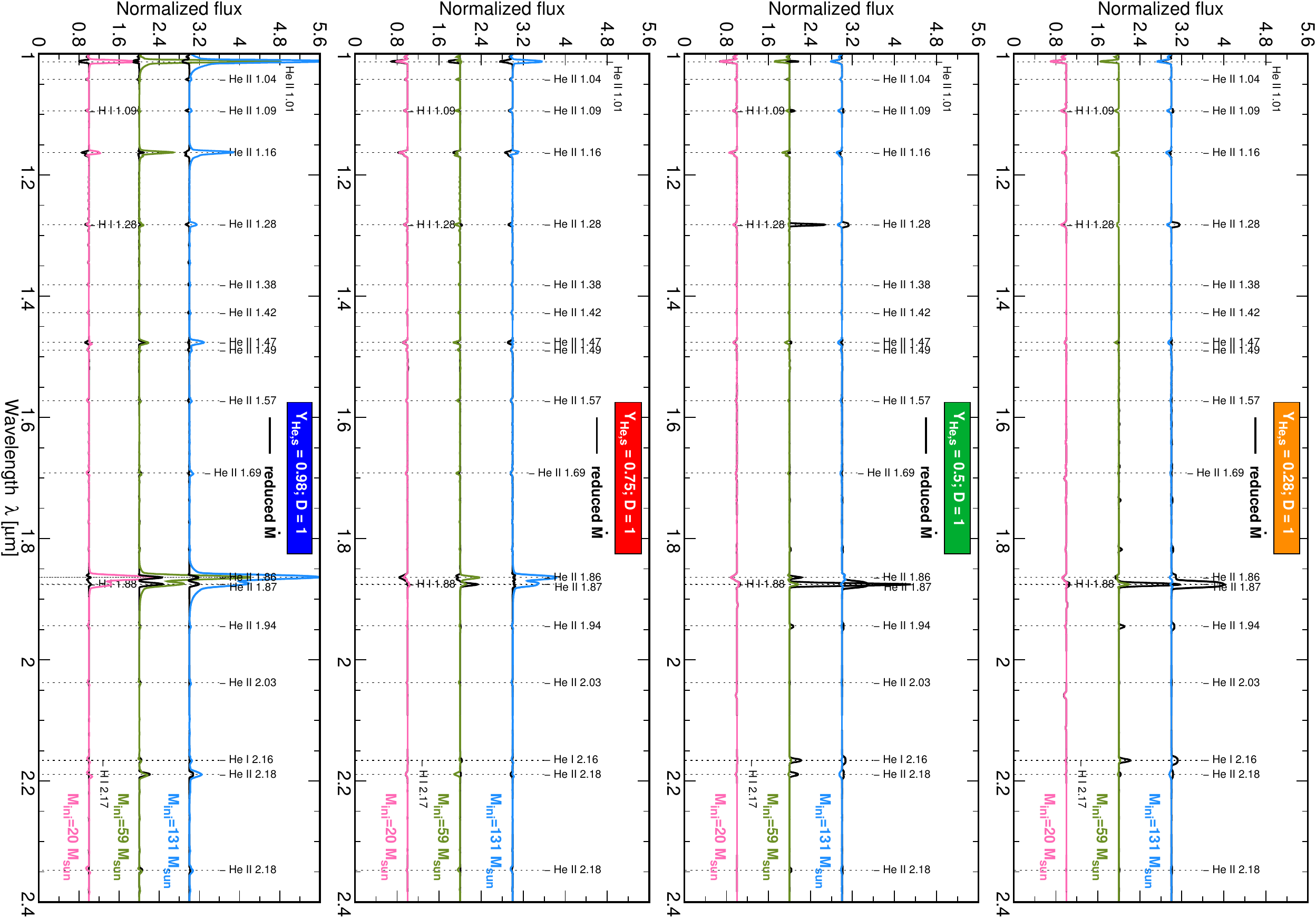}
			\caption{Same as Fig.~\ref{fig:MS-opt-D1} but in the IR region.}\label{fig:MS-ir-D1}
		\end{figure*}
		
		%!!!!!! pMS D = 1 and D = 10. !!!!!!!
		\begin{figure*}\centering
			\includegraphics[width=0.98\textwidth,angle=90]{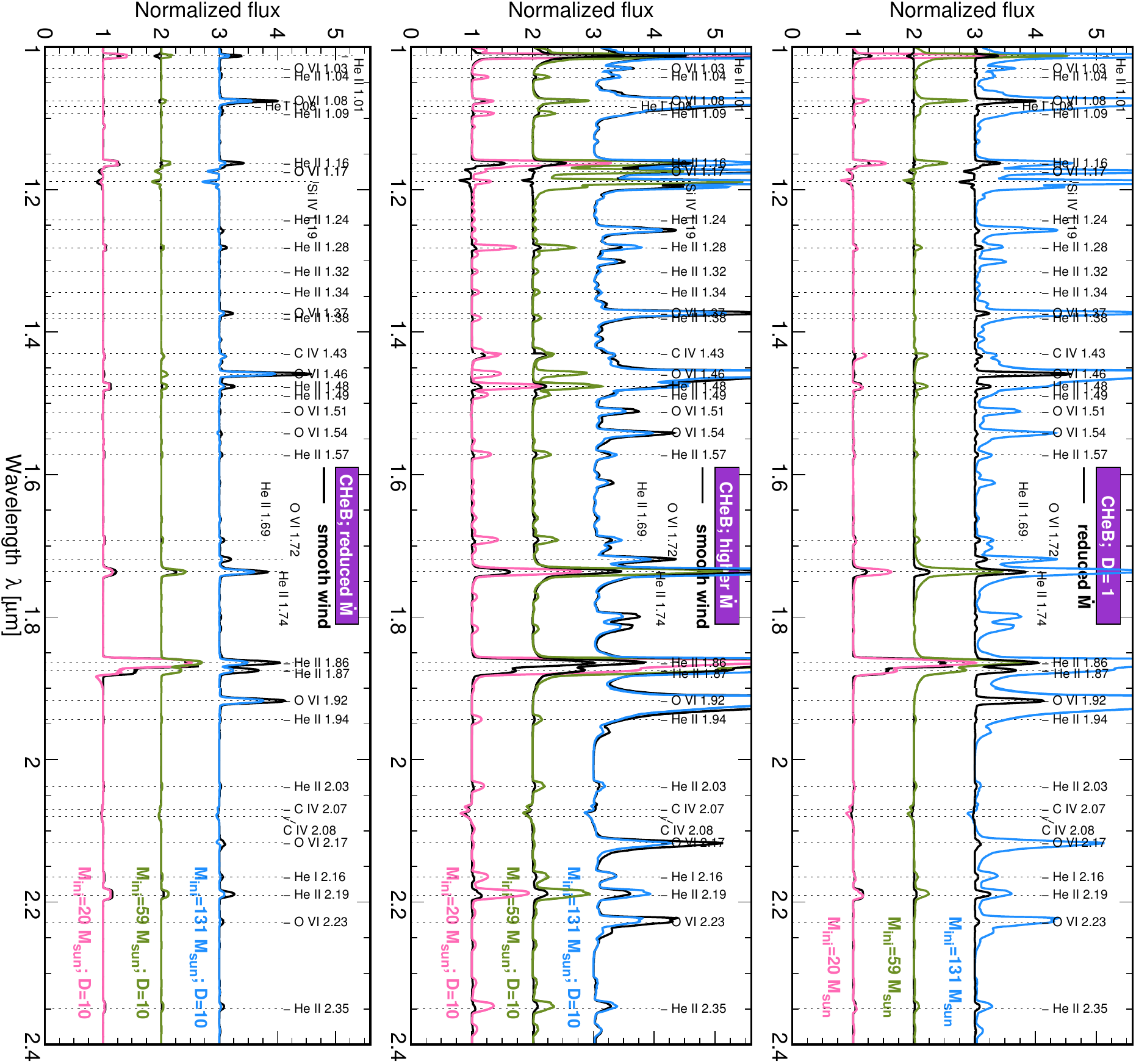}
			\caption{Same as Fig.~\ref{fig:pMS-opt-D1-D10-r} but in the IR region.}\label{fig:pMS-ir-D1-D10-r}
		\end{figure*}
		
		%%%%%%%  OPTICAL REGION  MS D = 10 %%%%%%%%%%%%%%%%%%%%%%%%%%%%%%%%%%%%%%%%%%%%%%%%%%%%%%
		\begin{figure*}[htp]\centering
			\includegraphics[width=1.3\textwidth,angle=90]{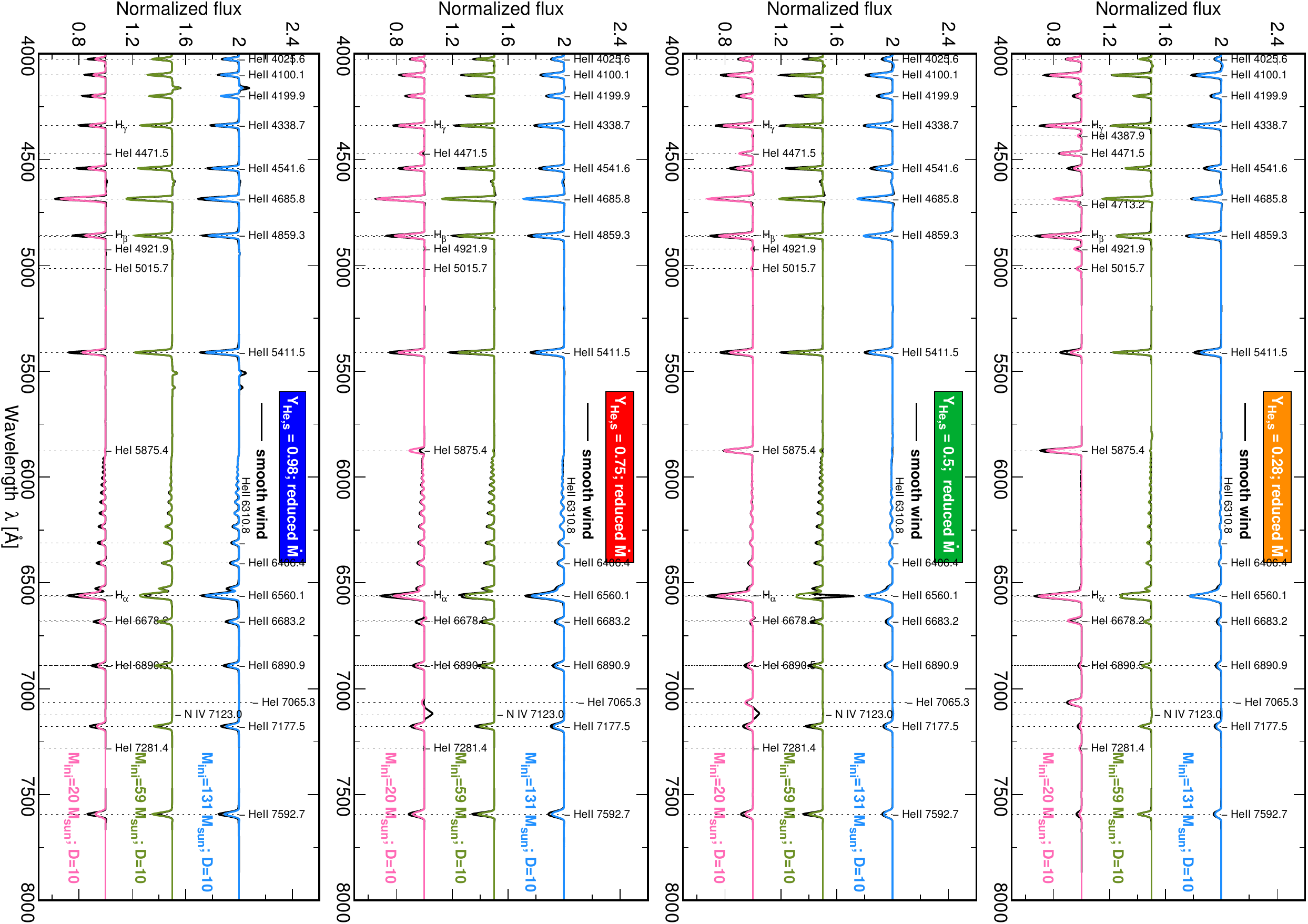}
			\caption{Same as Fig.~\ref{fig:MS-opt-D1-D10-hmd} but assuming a mass-loss rate hundred times lower than nominal (i.e. higher) value.}
			\label{fig:MS-opt-D1-D10-rmd}
		\end{figure*}
		
		%%%%%%%  UV REGION MS D = 10 %%%%%%%%%%%%%%%%%%%%%%%%%%%%%%%%%%%%%%%%%%%%%%%%%%
		
		\begin{figure*}\centering
			\includegraphics[width=1.3\textwidth,angle=90]{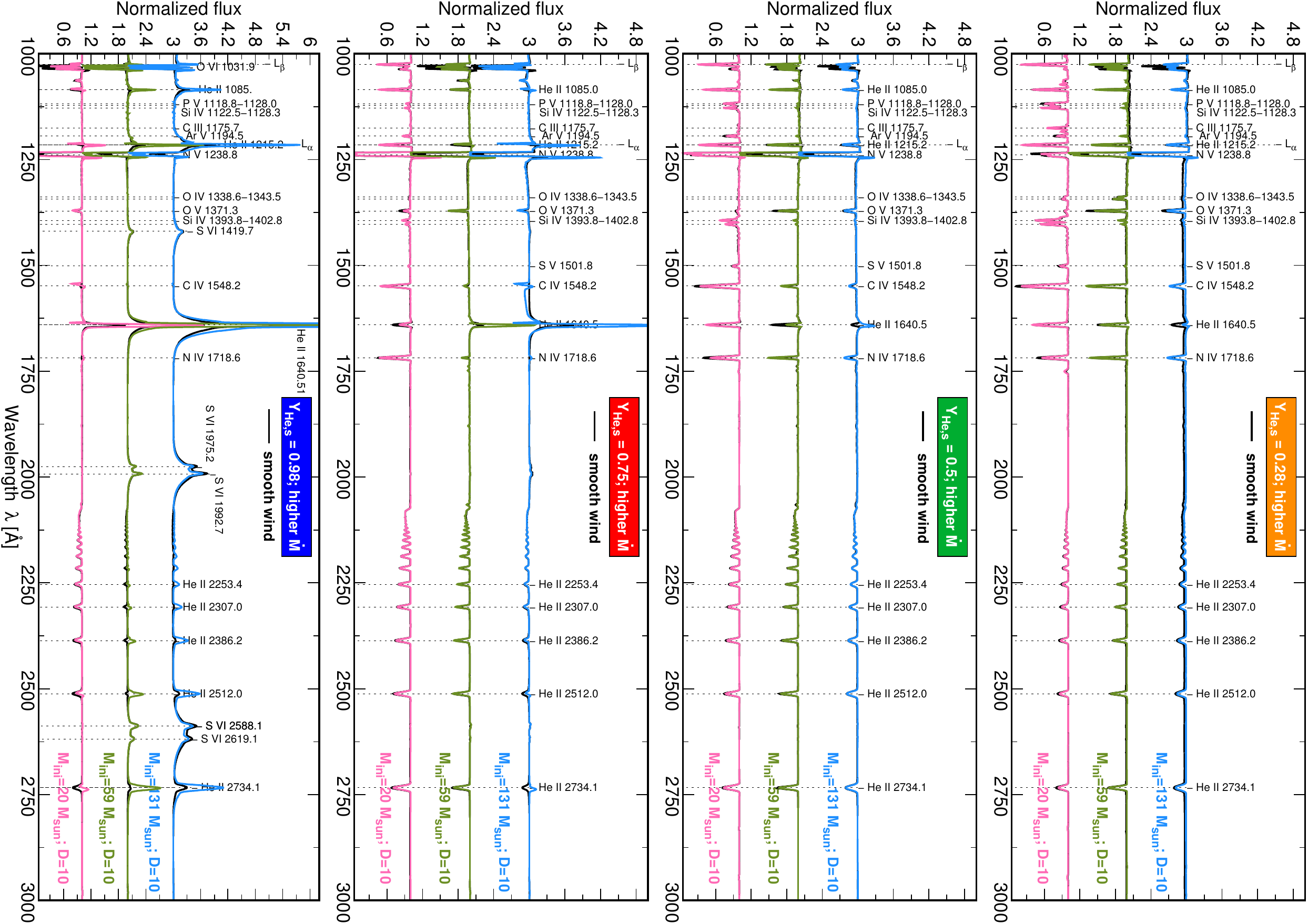}
			\caption{Same as Fig.~\ref{fig:MS-opt-D1-D10-hmd} but in the UV region.}\label{fig:MS-uv-D1-D10-hmd}
		\end{figure*}
		
		\begin{figure*}\centering
			\includegraphics[width=1.3\textwidth,angle=90]{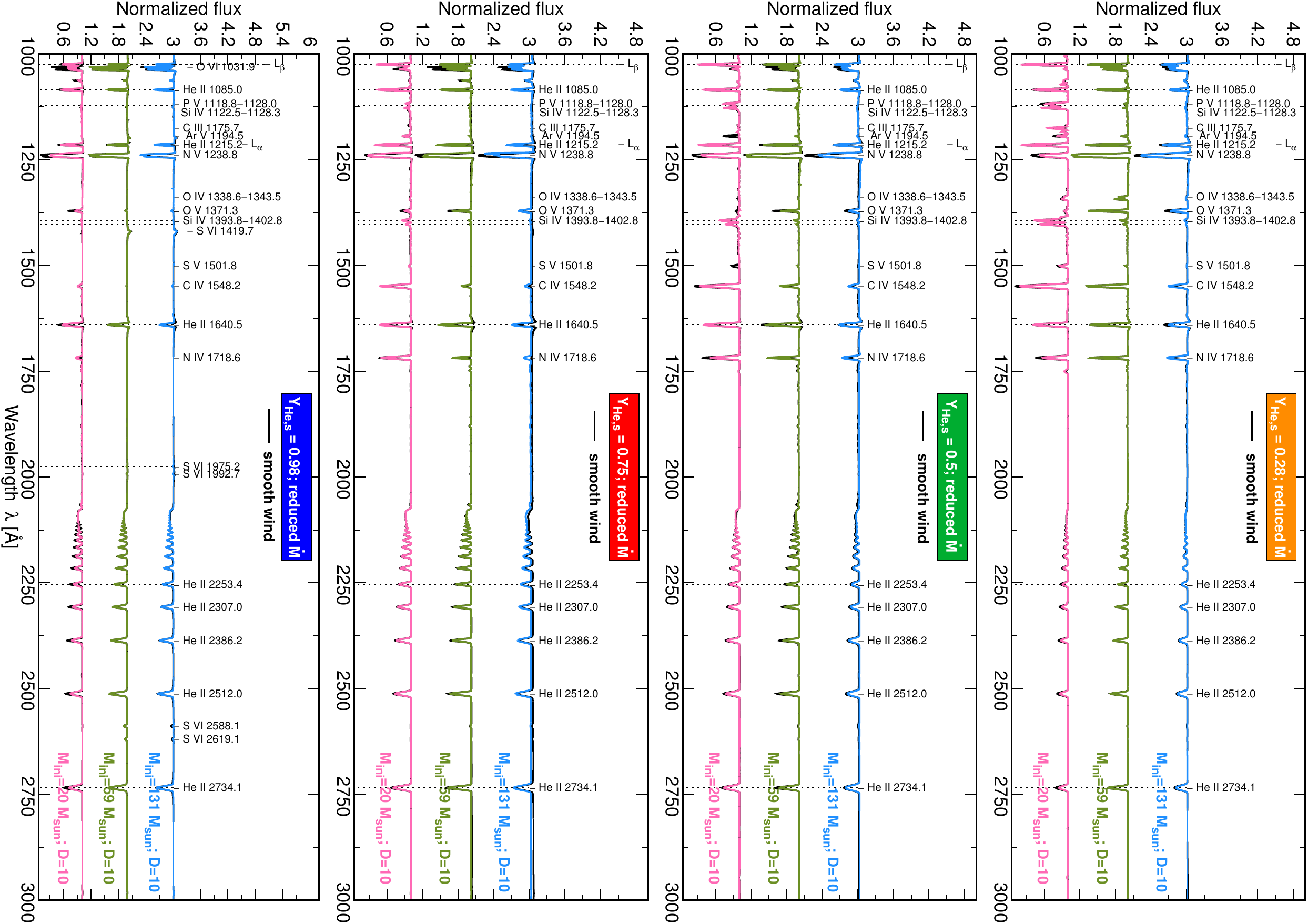}
			\caption{Same as Fig.~\ref{fig:MS-opt-D1-D10-rmd} but in the UV region.}\label{fig:MS-uv-D1-D10-rmd}
		\end{figure*}
		
		%%%%%%%  IR REGION MS D = 10 %%%%%%%%%%%%%%%%%%%%%%%%%%%%%%%%%%%%%%%%%%%%%%%%%%%
		
		\begin{figure*}\centering
			\includegraphics[width=1.3\textwidth,angle=90]{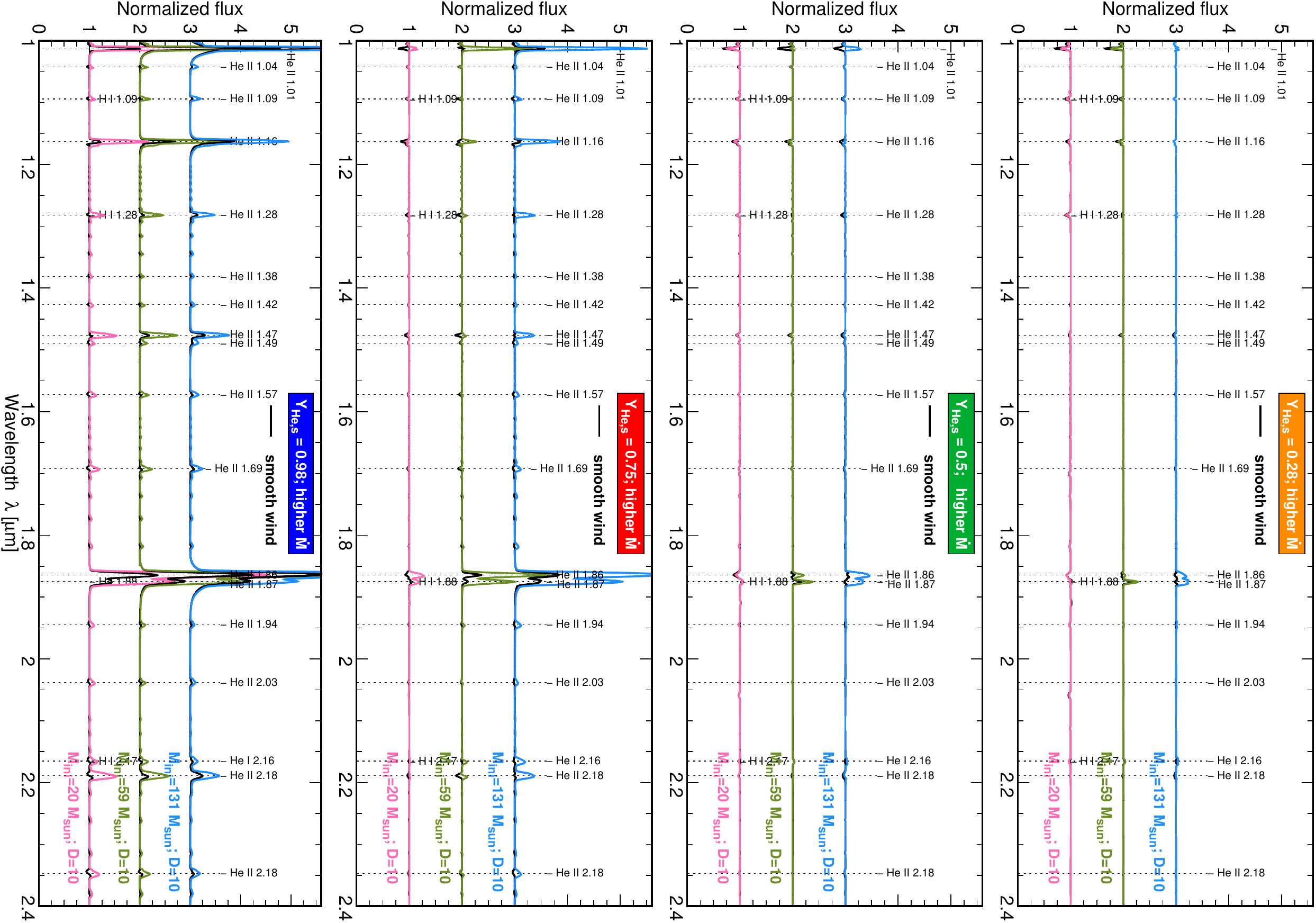}
			\caption{Same as Fig.~\ref{fig:MS-opt-D1-D10-hmd} but in the IR region.}\label{fig:MS-ir-D1-D10-hmd}
		\end{figure*}
		
		\begin{figure*}\centering
			\includegraphics[width=1.3\textwidth,angle=90]{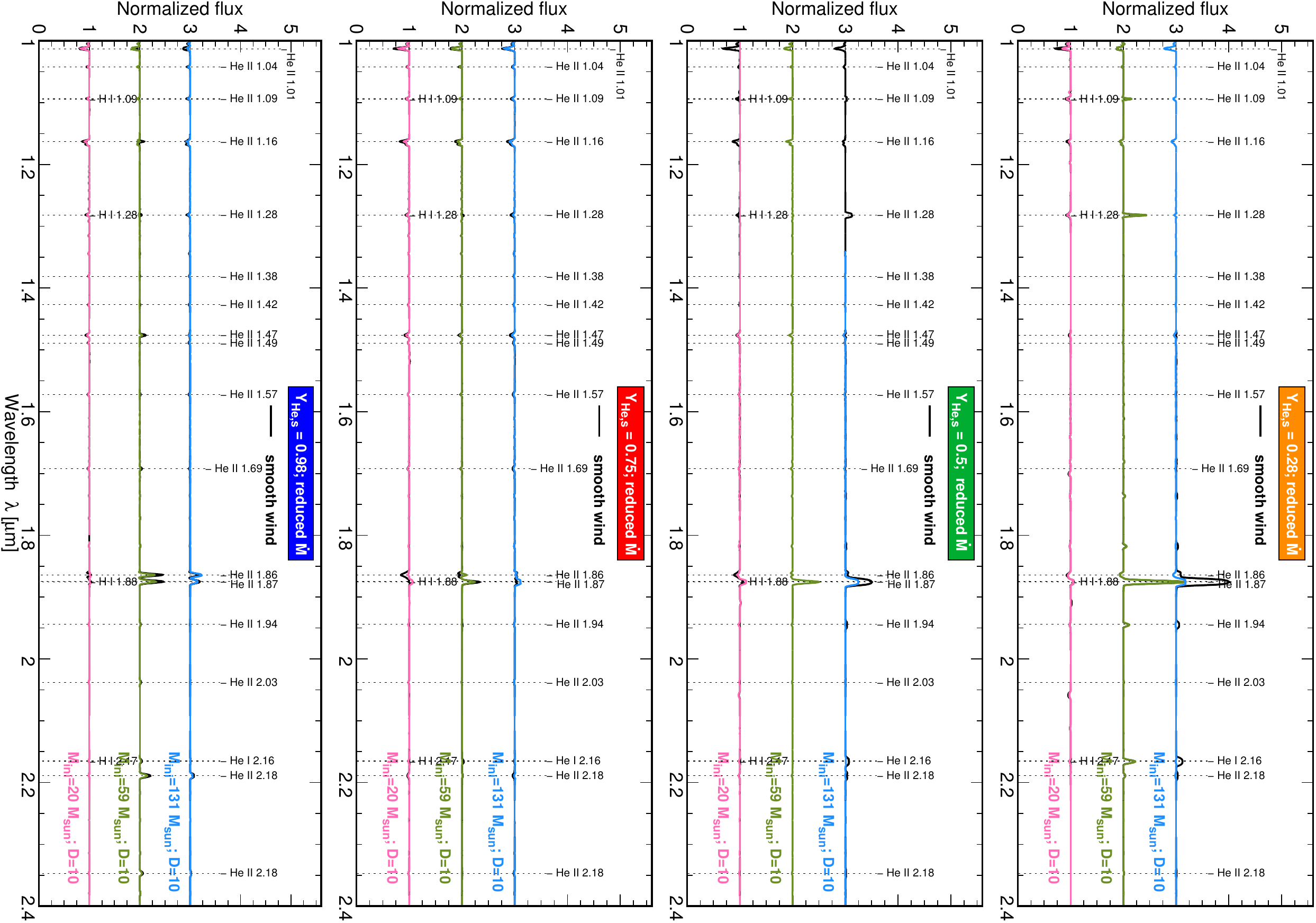}
			\caption{Same as Fig.~\ref{fig:MS-opt-D1-D10-rmd} but in the IR region.}\label{fig:MS-ir-D1-D10-rmd}
		\end{figure*}

		%%%%%%%%%%%%%%%%%%%%%%%%%%%%%%%%%%%%%%%%%%%%%%%%%%%%%%%%%%%%%%%%%%%%%%%%%%%%%%%%

	\end{appendix}
	
\end{document}